\documentclass[reprint, amsmath, amssymb, aps, nofootinbib]{revtex4-2}

\usepackage{graphicx}
\usepackage{dcolumn}
\usepackage{bm}
\usepackage{braket}
\usepackage{amsmath}
\usepackage{here}
\usepackage{color}
\usepackage{comment}

\usepackage[small]{caption}
\usepackage[subrefformat=parens]{subcaption}
\begin{document}
\title{
Dimensional reduction gauge and effective dimensional reduction \\
in the four-dimensional Yang-Mills theory
}
\author{Kei~Tohme}
\author{Hideo~Suganuma}
\affiliation{
Department of Physics, 
Kyoto University, \\ 
Kitashirakawaoiwake, Sakyo, Kyoto 606-8502, Japan}
\date{\today} 
\begin{abstract}
{
Motivated by one-dimensional color-electric flux-tube formation in 
four-dimensional (4D) QCD, we investigate a possibility of 
effective dimensional reduction 
in the 4D Yang-Mills (YM) theory.
%
We propose a new gauge fixing of ``dimensional reduction (DR) gauge" defined so as to minimize
$R_{\mathrm{DR}}~\equiv~\int d^{4}s ~ \mathrm{Tr} \left[ A_{x}^{2}(s) + A_{y}^{2}(s) \right]$,
which has 
a residual gauge symmetry for the gauge function $\Omega (t,z)$ like 2D QCD on the $t$-$z$ plane.
We investigate effective dimensional reduction 
in the DR gauge 
using SU(3) quenched lattice QCD at $\beta = 6.0$. 
The amplitude of $A_{x}(s)$ and $A_{y}(s)$ are found to be strongly suppressed in the DR gauge.
We consider ``$tz$-projection" of $A_{x,y}(s) \to 0$ 
for the gauge configuration generated in the DR gauge, in a similar sense to Abelian projection in the maximally Abelian gauge. 
By the $tz$-projection in the DR gauge, the interquark potential is not changed, and $A_{t}(s)$ and $A_{z}(s)$ play a dominant role in quark confinement.
In the DR gauge, we calculate a spatial correlation 
$\langle \mathrm{Tr} A_{\perp}(s) A_{\perp}(s+ra_{\perp}) \rangle ~ (\perp = x,y)$ and estimate the spatial mass of $A_{\perp}(s) ~ (\perp = x,y)$ 
as $M \simeq 1.7 ~ \mathrm{GeV}$. 
It is conjectured that this large mass makes $A_{\perp}(s)$ inactive and realizes the dominance of $A_{t}(s)$ and $A_{z}(s)$ in infrared region in the DR gauge.
We also calculate the spatial correlation of two temporal link-variables and find that the correlation decreases as $\exp (-mr)$ with $m \simeq 0.6 ~ \mathrm{GeV}$. 
Using a crude approximation, the 4D YM theory is reduced into an ensemble of 2D YM systems with the coupling of $g_{\rm 2D} = g m$.
}
\end{abstract}

\maketitle

\section{Introduction}
\label{sec:intro}

Quantum chromodynamics (QCD) is the fundamental theory of strong interactions.
However, its analytical solving is an important difficult problem even at present due to its strong-coupling nature in low-energy region \cite{PhysRevLett.30.1343, PhysRevLett.30.1346}. 
In particular, understanding the mechanism of quark confinement is one of the most difficult problems in four-dimensional (4D) QCD and has not yet been solved by analytical methods. 

In 4D QCD, quark confinement is characterized by a linear interquark potential  
and one-dimensional squeezing of color-electric fields, which is idealized as the string picture in the infrared region. Here, the string tension $\sigma \simeq 0.89 {\rm GeV/fm}$ gives the quark confining force and is the key parameter of quark confinement.
Historically, the string picture of hadrons was proposed to explain 
the Regge trajectory of hadrons \cite{NAMBU_1969,GOTO_1971}, and the interquark potential was investigated from the spectra of heavy quarkonia \cite{Cornell}. 

After lattice QCD was performed as the first principle calculations of the strong interaction \cite{CREUTZ_MCSU2_1980}, many lattice QCD studies have shown that the interquark potential is expressed to be a sum of the one-gluon-exchange Coulomb part and a linear part belonging to the string picture \cite{Takahashi_Suganuma_detailedQQpot_2002,Rothe_LQCDtext,Handbook_2023}.
The one-dimensional color-flux-tube formation is also directly observed between (anti)quarks in lattice QCD calculations \cite{Rothe_LQCDtext, ICHIE2003C899, kitazawa}.

In nature, ordinary waves propagate isotropically.
In fact, electromagnetic fluxes, gravitational waves and sound waves spread over three-dimensional space.
In contrast, as schematically shown in Fig.1,
a color-electric flux is squeezed one-dimensionally in QCD, which can be regarded as a reduction of the spatial dimension by two. 
The one-dimensional flux-tube formation might be considered as a kind of effective dimensional reduction. 
Note also that the flux-tube formation explains not only quark confinement but also gluon confinement because the gluonic flux is confined inside a narrow tube area between (anti)quarks, unlike the widely spread electromagnetic flux in QED \cite{Handbook_2023}.

\begin{figure}[h]
    \centering
    \includegraphics[width=8.0cm]{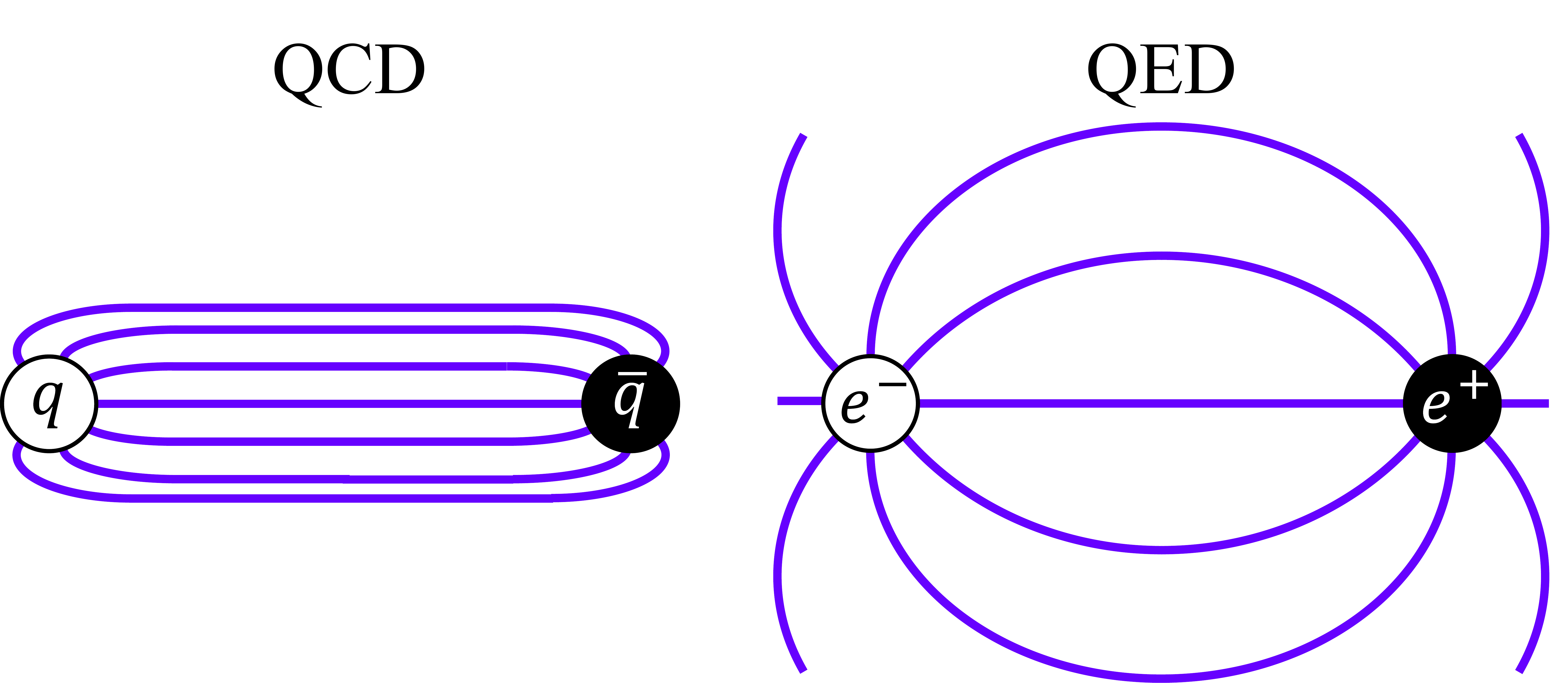}
\caption{
    Schematic figure of electric-fluxes in QCD and QED.
    The left represents the one-dimensional flux-tube formation in QCD, and the right the flux in QED which spreads over three-dimensional space.
    }
    \label{fig:flux_compare}
\end{figure}

Such a flux-squeezing is also observed as the Abrikosov vortex in type-I\hspace{-1.2pt}I superconductors immersed in magnetic fields, where the Meissner effect induced by Cooper-pair condensation repels the magnetic fields and a one-dimensional magnetic-flux-tube is formed.

Motivated by the Abrikosov vortex, Nambu, 't~Hooft and Mandelstam proposed the dual superconductor picture \cite{NAMBU_DSC_1974, HOOFT_DSC_1976, MANDELSTAM_DSC_1976} to explain the color-flux-tube formation in 4D QCD.
In this picture, color-electric flux is squeezed by the dual Meissner effect due to color-magnetic-monopole condensation. 
Although QCD does not contain monopoles explicitly, they appear as topological objects in the Abelian gauge \cite{HOOFT1981455}.
Using maximally Abelian (MA) gauge, the dual superconductor picture has been demonstrated in lattice QCD \cite{KRONFELD_MAG2_1987, SUZUKI_YOTSUYANAGI_AD, Brandstaeter_MonopoleCond_1991, Miyamura_MonopoleChralCond_1995, Woloshyn_MonopoleChralCond_1995, Chernodub_Polikarpov_1997, 
Ichie_Suganuma_1999, Amemiya_Suganuma_1999, Gongyo_Iritani_Suganuma_2012,
SAKUMICHI_SUGANUMA_AD, 
Sakumichi_Suganuma_AD_2015, Ohata_Suganuma_MonopoleChralCond_2021}.

As another example of effective dimensional reduction, the Parisi-Sourlas mechanism \cite{Parisi_Sourlas_1979} shows the equivalence of a $d$-dimensional spin system under Gaussian random magnetic fields and the $(d-2)$-dimensional system without magnetic fields.

As for the QCD vacuum, it is pointed out that the $\mathrm{SU(2)}$ Yang-Mills theory 
has color-magnetic instability, and color-magnetic fields are spontaneously generated, which is called the Savvidy vacuum \cite{Savvidy_1977}.
In fact, the gluon condensate is found to be positive,  
\begin{equation}
    \label{eq:gluon_cond}
    \left\langle
        \frac{\alpha_{s}}{\pi}
        G^{a}_{\mu \nu} G^{a \, \mu \nu}
    \right\rangle
    \propto
    \left\langle
        H^{2}_{a}-E^{2}_{a}
    \right\rangle
    > 0, 
\end{equation}
in the QCD sum rule \cite{Shifman_QCDsum_1979} 
and 
also in lattice QCD \cite{Giacomo_Rossi_GluonCond_1981, Kripfganz_GluonCond_1981, Ilgenfritz_Müller_GluonCond_SU3_1982}. 
Thus, the QCD vacuum is considered to be filled with color-magnetic fields.
Considering that the ground-state solution is not uniform at the one-loop level \cite{Nielsen_Olesen_1979}, 
the real QCD vacuum is conjectured to be filled with random color-magnetic fields at a large scale, which is called the Copenhagen (spaghetti) vacuum  \cite{Ambjorn_Olesen_1980}.
There might be some connection between the effective dimensional reduction in 4D QCD and the Parisi-Sourlas mechanism, where random magnetic fields play an important role \cite{Iritani_Suganuma_Iida_2009}.

In any case, ``effective dimensional reduction'' might be a key concept in 4D QCD.
In this paper, we consider a way to show 
effective dimensional reduction in 4D QCD 
and demonstrate it in the 4D Yang-Mills (YM) theory, 
focusing gauge degrees of freedom. 

In particular, we investigate the possibility of describing 
the 4D YM theory in terms of 2D YM-like degrees of freedom.
Such a description has some merits, and one of them is its analyticity.
In 2D QCD, analytical methods are more effective than in 4D QCD, and the meson description is performed in the large $N_{c}$ limit \cite{HOOFT_2dMeson_1974}.
Another merit is that, in two-dimensional spacetime, even the tree-level potential is linear, and quark confinement is automatically realized.

To clarify the low-dimensional picture in 4D QCD, 
we utilize gauge degrees of freedom, 
and propose a new gauge fixing of 
``dimensional reduction (DR) gauge''.
Note here that, 
while physical quantities are gauge invariant, 
a physical picture could be gauge dependent 
like the dual superconductor picture in the MA gauge.
In fact, there might be a suitable gauge to 
see effective dimensional reduction in 4D QCD. 


Now, we mention ``utility of gauge fixing'' to obtain a mathematical description or a semi-physical picture for some phenomenon. 
One of the merit to use a specific gauge fixing is to find out a semi-physical picture to grasp the phenomenon. 
Of course, all the physical quantities must be gauge invariant  
and can be described with the gauge-invariant quantities in principle. 
Nevertheless, there could be suitable gauge fixing 
to get a mathematical description or a semi-physical picture 
based on some physical analogy.

One typical example is the use of MA gauge fixing  
for the Abelian dual superconductor picture
\cite{KRONFELD_MAG2_1987, SUZUKI_YOTSUYANAGI_AD, Brandstaeter_MonopoleCond_1991, Miyamura_MonopoleChralCond_1995, Woloshyn_MonopoleChralCond_1995, Chernodub_Polikarpov_1997, 
Ichie_Suganuma_1999, Amemiya_Suganuma_1999, Gongyo_Iritani_Suganuma_2012,
SAKUMICHI_SUGANUMA_AD, 
Sakumichi_Suganuma_AD_2015, Ohata_Suganuma_MonopoleChralCond_2021}, 
where the electromagnetic duality is manifest. 
In the MA gauge, the YM theory is described as an Abelian gauge theory 
with Abelian gauge fields and adjoint charged matter fields. 
This description is suitable for the Abelian dual superconductor picture. 
Even without gauge fixing, the similar mechanism might be embedded in 
the YM theory. However, the way of the description is highly nontrivial and complicated without use of the MA gauge.

The other example is the use of covariant gauge fixing 
to find the Kugo-Ojima criterion \cite{Kugo-Ojima_1979} for color confinement, 
which is sometimes called the inverse Higgs-mechanism theorem, 
and also to get the Gribov-Zwanziger horizon condition \cite{Gribov_1978,Zwanziger89,Zwanziger92,
Zwanziger93}.
To get the Kugo-Ojima-Gribov-Zwanziger picture, one has to take 
a globally SU($N_c$)-symmetric covariant gauge such as the Landau gauge, 
since the BRST and Lorentz symmetry play the key role to formulate.
This confinement picture has been investigated also 
in lattice QCD in the Landau gauge \cite{Furui-Nakajima_2004}.

A familiar example is the use of the Coulomb gauge 
for canonical quantization in QED 
because the canonical structure becomes clear \cite{H92}
in the Coulomb gauge $\nabla \cdot A = 0$. 
Although Lorentz invariance is apparently broken in the Coulomb gauge, it is recovered when physical quantities are calculated.

In this way, in spite of gauge invariance of physical quantities, 
some specific gauge fixing could be economically useful 
to get a semi-physical picture or a mathematical description for each physical phenomenon.

This paper is organized as follows.
In Sec.\ref{sec:fromulation_continuum}, we formulate the DR gauge and $tz$-projection in continuous QCD.
We also formulate them in lattice QCD in Sec.\ref{sec:DR_LQCD}.
In Sec.\ref{sec:numerical}, we perform the lattice QCD calculations and numerical analyses in the DR gauge.
In Sec.\ref{sec:discussion}, we discuss on analytical modeling of the YM theory 
in the DR gauge with an approximation.
Section \ref{sec:summary} is devoted for the summary and concluding remarks.

\section{Dimensional Reduction gauge: formulation in continuum QCD}
\label{sec:fromulation_continuum}
In this section, we define ``dimensional reduction (DR) gauge'' and ``$tz$-projection'' in continuous QCD.
In this paper, we use $s = (x,y,z,t)$ as a space-time coordinate four-vector.

The $\mathrm{SU}(N_{c})$ QCD action is given as
\begin{equation}
    \label{eq:action_QCD}
    \hspace{-5pt}
    S_{\mathrm{QCD}} = 
    \int d^{4} s \;
    \left[ 
    -
    \frac{1}{2}
    \mathrm{Tr} \; G_{\mu\nu}G^{\mu\nu}
    +
    \Bar{q}
    \left(
    i \gamma^{\mu} D_{\mu} - m 
    \right)
    q
    \right],
\end{equation}
where $q$ is quark field and $m$ a current quark mass.
The covariant derivative $D_{\mu}$ is defined by $\mathrm{SU}(N_{c})$ gluon field $A_{\mu} \in \mathfrak{su}(N_{c})$ and the QCD gauge coupling $g$ as
\begin{equation}
    \label{eq:cov_der}
    D_{\mu}
    \equiv
    \partial 
    +
    i g A_{\mu},
\end{equation}
and the field strength tensor $G_{\mu \nu}$ is defined as
\begin{equation}
    \label{eq:str_tensor}
    G_{\mu \nu}
    \equiv
    \frac{1}{ig} \left[ D_{\mu}, D_{\nu} \right]
    =
    \partial_{\mu} A_{\nu} - \partial_{\nu} A_{\mu}
    + ig \left[ A_{\mu}, A_{\nu} \right].
\end{equation}
In this study, we only deal with the gauge part of the action \eqref{eq:action_QCD},
that is, the YM theory.

\subsection{Definition of dimensional reduction (DR) gauge}
\label{subsec:DRG}
Dimensional reduction (DR) gauge is defined so as to minimize
\begin{eqnarray}
    \label{eq:DRG}
    R_{\mathrm{DR}}
    & \equiv &
    \int d^{4} s \hspace{-5pt}
    \sum_{\perp = x, y} \hspace{-5pt}
    \mathrm{Tr}\left[ 
    A_{\perp}(s)^2
    \right] 
    \nonumber \\
    & = &
    \int \hspace{-2pt}
    d^{4} s \;
    \mathrm{Tr}\left[ 
    A_{x}(s)^2 + A_{y}(s)^2
    \right]
\end{eqnarray}
with the gauge transformation.
Here, the subscript $\perp$ denotes $x$ and $y$ 
in this paper.
Since $R_{\mathrm{DR}}$ does not contain $A_{t}(s)$, the DR gauge can be defined in Minkowski spacetime, and this gauge fixing can be perform locally in the temporal direction, like the Coulomb gauge.

The DR gauge has a residual gauge symmetry for the gauge function $\Omega(t,z)$.
In fact, with the gauge function $\Omega(t,z)$, the gauge fields 
transform as
\begin{eqnarray}
    \label{eq:res_sym_tz}
    A_{t,z}(s) 
    & \to &
    \Omega(t,z)
    \left( 
    A_{t,z}(s) + \frac{1}{ig}\partial_{t,z}
    \right)
    \Omega^{\dagger}(t,z) , \\
    \label{eq:res_sym_xy}
    A_{\perp}(s) 
    & \to &
    \Omega(t,z)
    \left( 
    A_{\perp}(s) + \frac{1}{ig}\partial_{\perp}
    \right)
    \Omega^{\dagger}(t,z) \nonumber \\
    & = & 
    \Omega(t,z)
    A_{\perp}(s)
    \Omega^{\dagger}(t,z) .
\end{eqnarray}
Since 
$R_{\mathrm{DR}}$ in Eq.\eqref{eq:DRG} is invariant under this partial gauge transformation, DR-gauged QCD has the residual symmetry.

Note that this residual gauge symmetry is the same as 2D QCD on the $t$-$z$ plane.
From the gauge transformation \eqref{eq:res_sym_tz}, $A_{t}(s)$ and $A_{z}(s)$ correspond to the gauge fields in 2D QCD. 
On the other hand, the gauge transformation \eqref{eq:res_sym_xy} represents that $A_{x}(s)$ and $A_{y}(s)$ can be interpreted as charged matter belonging to the adjoint representation in 2D $\mathrm{SU}(N_{c})$ gauge theory.
Thus, DR-gauged QCD can be regarded as 2D $\mathrm{SU}(N_{c})$ gauge theory with gauge fields $A_{t,z}(s)$ and charged matter fields $A_{x,y}(s)$.

The above definition of the DR gauge is a global definition.
The local condition of the DR gauge is given by 
\begin{eqnarray}
    \label{eq:DRG_local}
    \sum_{\perp = x, y}
    \partial_{\perp}A_{\perp}(s) 
    \equiv 
    \partial_{x}A_{x}(s) + \partial_{y}A_{y}(s) = 0 .
\end{eqnarray}
similar to the Landau gauge or the Coulomb gauge.
This local condition is derived as the minimal condition of Eq.\eqref{eq:DRG} with gauge transformation.

Including the gauge fixing term,  
the gauge action of DR-gauged QCD is expressed as 
\begin{equation}
    \label{eq:DRG_action}
    S_{\mathrm{DR}} = 
    \int d^{4} s 
    \left[ 
    -
    \frac{1}{2}
    \mathrm{Tr} \; G_{\mu\nu}G^{\mu\nu}
    +
    \frac{1}{2\alpha}
    \sum_{\perp = x, y}
    \mathrm{Tr}
    \left( \partial_{\perp}A_{\perp} \right)^{2}
    \right] \! .
\end{equation}
The second term is for the DR gauge fixing ($\alpha$ is the gauge fixing parameter), 
and it 
is invariant for the residual gauge transformation \eqref{eq:res_sym_xy} as
\begin{eqnarray}
    \label{eq:gt_res_gfix}
    \mathrm{Tr} 
    \left( \partial_{\perp}A_{\perp} \right)^{2}
    && \to
    \mathrm{Tr} 
    \left\{
        \Omega(t,z)
        \partial_{\perp}A_{\perp}
        \Omega^{\dagger}(t,z)
    \right\}^{2}
    \nonumber \\
    && =
    \mathrm{Tr} 
        \left(
        \partial_{\perp}A_{\perp}
        \right)^{2}.
\end{eqnarray}
Therefore, DR-gauged QCD has a residual gauge symmetry for $\Omega(t,z)$, and it is the same as that 2D QCD on the $t$-$z$ plane.

In the path-integral formalism in the Euclidean metric, 
the generating functional of DR-gauged QCD is expressed as 
\begin{eqnarray}
Z&=&\int \mathcal{D}A \mathcal{D}q \mathcal{D}{\bar q} e^{-S_{\rm QCD}} \delta(\partial_\perp A_\perp) {\rm Det}(\partial_\perp D_\perp) \\
&=&\int \mathcal{D}A \mathcal{D}q \mathcal{D}{\bar q} \mathcal{D}c \mathcal{D}{\bar c}~e^{-S_{\rm QCD}+\int d^4s~{\bar c}(\partial_\perp D_\perp)c} \delta(\partial_\perp A_\perp), \nonumber
\end{eqnarray}
where the last factor in the first line is the Faddeev-Popov (FP) determinant \cite{Faddeev_Popov_1967}
and $c$ and ${\bar c}$ denote the FP ghost and anti-ghost fields, respectively.
Note that the additional factors relating to the DR-gauge fixing 
does not include $A_t$ and $A_z$, and  
the action form on $A_{t,z}$ is formally unchanged. 
Accordingly, the FP (anti-)ghost field directly couples only to $A_\perp$ in the DR gauge.

Then, the generating functional $Z$ is formally rewritten as 
\begin{eqnarray}
Z&=&\int \mathcal{D}A_{t,z} \mathcal{D}A_\perp \mathcal{D}q \mathcal{D}{\bar q}~ 
e^{-S_{t,z}[A_{t,z}]} e^{-S_{\rm cross}[A_{t,z},A_\perp]} \cr
&&~~~~~~~~~~e^{-S_\perp[A_\perp]}\delta(\partial_\perp A_\perp) {\rm Det}(\partial_\perp D_\perp) \cr
&=&\int \mathcal{D}A_{t,z} \mathcal{D}A_\perp \mathcal{D}q \mathcal{D}{\bar q}~ 
e^{-S_{t,z}[A_{t,z}]} e^{-S_{\rm cross}[A_{t,z},A_\perp]} \cr
&&~~~~~~~~~~e^{-S_{\rm eff}[A_\perp]},
\label{eq:PI_DR}
\end{eqnarray}
where $S_{t,z}[A_{t,z}]$ and $S_\perp[A_\perp]$ are defined by
\begin{eqnarray}
S_{t,z}[A_{t,z}]&=&S_{\rm QCD}[A_{t,z}; A_\perp \equiv 0], \cr 
S_\perp[A_\perp]&=&S_{\rm QCD}[A_\perp; A_{t,z} \equiv 0], 
\end{eqnarray}
and $S_{\rm cross}[A_{t,z},A_\perp]$ denotes the cross term including both $A_{t,z}$ and $A_\perp$.
For the simple notation, the quark field has been abbreviated in the argument of the actions.
The second line of Eq.(\ref{eq:PI_DR}) only includes $A_\perp$, 
and then the action/interaction form of $A_{t,z}$ is unchanged in $Z$, while 
the effective action $S_{\rm eff}[A_\perp]$ can be drastically changed from the original one. 

As a general caution, similarly in the Landau, Coulomb and MA gauges, the DR gauge fixing also has 
the Gribov ambiguity \cite{Gribov_1978}, 
which is a general problem appearing in gauge fixing, 
except for the axial-gauge-type fixing.

\subsection{$tz$-projection}

\label{subsec:DRQCD}

In the DR gauge, 
from its definition with Eq.\eqref{eq:DRG}, 
the amplitudes of $\perp$-directed gluon fields $A_{x}(s)$ and $A_{y}(s)$ would be strongly suppressed.
To investigate the possibility of describing 4D QCD in terms of 2D QCD-like degrees of freedom, 
we introduce ``$tz$-projection'' 
as removal of the $\perp$-directed gluon fields $A_{x}(s)$ and $A_{y}(s)$, i.e., 
the replacement of 
\begin{eqnarray}
    \label{eq:tz_proj}
    A_{x,y}(s) \to 0.
\end{eqnarray}
Then, applying the $tz$-projection in the DR gauge, 
we investigate properties on effective dimensional reduction  
in the 4D YM theory.

\begin{figure}[h]
\begin{center}
    \includegraphics[width=6.0cm]{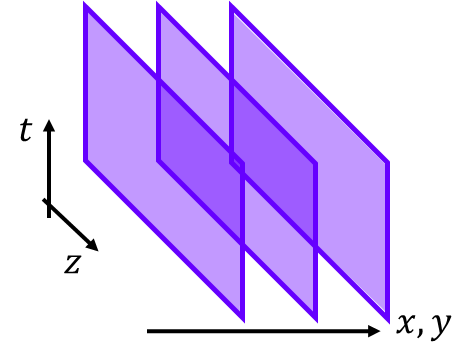}
\end{center}
\caption{
    Schematic figure of the DR-gauged YM theory 
    under the $tz$-projection.
    2D YM-like systems are piled in the $x$ and $y$ directions, and they interact with the piled neighbors.
    }
    \label{fig:tz_proj.ed_QCD4_4d}
\end{figure}

After the $tz$-projection \eqref{eq:tz_proj}, 
the gauge action \eqref{eq:DRG_action} of DR-gauged QCD becomes
\begin{eqnarray}
    \label{eq:DRG_act_tzp}
    S^{tz}_{\mathrm{DR}} 
    = 
    \int dx dy
    && \int dt dz \;
    \Bigg[ 
    \mathrm{Tr} \; G^{2}_{tz} 
    \nonumber \\
    && 
    \hspace{-10pt}
    + \!
    \left.
    \sum_{\perp=x,y}
    \mathrm{Tr} 
    \left\{
      \left( \partial_{\perp}A_{t} \right)^2
    - \left( \partial_{\perp}A_{z} \right)^2
    \right\}
    \right]
\end{eqnarray}
at the tree-level.
In this action, the first term is equal to 
the 2D YM action on the $t$-$z$ plane.
The second term is interpreted as interaction between neighboring 2D YM-like systems in the $x$ and $y$ directions.
Thus, as shown in Fig.\ref{fig:tz_proj.ed_QCD4_4d}, 
the DR-gauged YM theory after the $tz$-projection can be expressed as 2D YM-like systems on $t$-$z$ planes piled in the $x$ and $y$ directions and these 2D systems interacting with each other.
In fact, the integration over $x$ and $y$ in Eq.\eqref{eq:DRG_act_tzp} represents that the 2D YM-like systems are piled in the $x$ and $y$ directions.
 
\section{Lattice Formalism of DR gauge}
\label{sec:DR_LQCD}
We formulate $\mathrm{SU}(N_{c})$ lattice QCD on a 4D lattice with spacing $a$ in the Euclidean spacetime.
In lattice QCD, the gauge degree of freedom is described as the link-variable $U_{\mu}(s) \equiv e^{iag A_{\mu}(s)} \in \mathrm{SU}(N_{c})$, instead of the gluon field $A_{\mu}(s) \in \mathfrak{su}(N_{c})$.
As the lattice gauge action, we use the standard plaquette action
\begin{equation}
\label{eq:plaquette}
    S^{\mathrm{lat}} 
    \equiv 
    \beta \sum_{s} \left[
    1 - \frac{1}{N_{c}}\sum_{\mu < \nu} \mathrm{ReTr} \; P_{\mu\nu}(s)
    \right],
\end{equation}
where $P_{\mu \nu}(s)$ denotes the plaquette variable defined as
\begin{equation}
    \label{eq:plaq}
    P_{\mu\nu}(s)
    \equiv
    U_{\mu}(s)
    U_{\nu}(s+a_{\mu})
    U^{\dagger}_{\mu}(s+a_{\nu})
    U^{\dagger}_{\mu}(s).
\end{equation}
Here, $a_{\mu}$ denotes the four vector in the $\mu$ direction with length of $a$.

\subsection{DR gauge and $tz$-projection in lattice QCD}
\label{subsec:DRG_lat}
In this subsection, 
we formulate the DR gauge and $tz$-projection in lattice QCD.
The DR gauge on a lattice is defined so as to maximize 
\begin{eqnarray}
    \label{eq:DRG_lat}
    R^{\mathrm{lat}}_{\mathrm{DR}}[U_\perp]
    & \equiv &
    \sum_{s}
    \sum_{\perp = x, y}
    \mathrm{Re Tr}
    \left[
        U_{\perp}(s)
    \right] \nonumber \\
    & = &
    \sum_{s}
    \mathrm{Re Tr}
    \left[
        U_{x}(s) + U_{y}(s)
    \right]
\label{eq:RDRlat}
\end{eqnarray}
with gauge transformation.
This corresponds to the definition with Eq.\eqref{eq:DRG} in the continuous limit, $a \to 0$.
In fact,
considering enough small $a$, $U_{\perp}(s)$ can be expanded as
\begin{equation}
    \label{eq:continuous_link}
    U_{\perp}(s)
    =
    1 + iag A_{\perp}(s) - \frac{1}{2} a^{2} g^{2} A^{2}_{\perp}(s)
    + O(a^{3}) ,
\end{equation}
and Eq.\eqref{eq:DRG_lat} 
becomes
\begin{equation}
    \label{eq:DRG_lat_a0}
    R^{\mathrm{lat}}_{\mathrm{DR}}
    \simeq
    - \frac{1}{2} a^{2} g^{2}
    \sum_{s}
    \sum_{\perp = x, y}
    \mathrm{Re Tr}
    \left[
        A_{\perp}(s)
    \right] 
\end{equation}
up to $O(a^{2})$, apart from a real constant.
Therefore, the maximization of $R^{\mathrm{lat}}_{\mathrm{DR}}$ corresponds to the minimization of $R_{\mathrm{DR}}$ in continuous spacetime.

\textcolor{black}{
Next, we derive the local condition of the DR gauge on a lattice.
Considering gauge transformation with $\Omega(s)$ only at a spacetime $s$, the variation $\delta R^{\mathrm{lat}}_{\mathrm{DR}}$ of the function $R^{\mathrm{lat}}_{\mathrm{DR}}$ is 
\begin{eqnarray}
    \label{eq:gt_lat_infinitesimal}
    \delta R^{\mathrm{lat}}_{\mathrm{DR}}
    & = &
    \sum_{\perp = x,y}
    \mathrm{Re Tr}
    \left[
        \Omega(s) U_{\perp}(s) + U_{\perp}(s-a_{\perp}) \Omega^{\dagger}(s)
    \right] \nonumber \\
    &&
    - 
    [
    \Omega(s)=1
    ]
    \nonumber \\
    & = &
    \mathrm{Re Tr}
    \left[
    \Omega(s)
    \sum_{\perp = x,y}
        \left\{
            U_{\perp}(s) + U^{\dagger}_{\perp}(s-a_{\perp})
        \right\}
    \right] \nonumber \\
    &&
    -
    [
    \Omega(s)=1
    ],
\end{eqnarray}
where $a_{\perp}$ is a four-vector in the $\perp$ direction with a length of lattice spacing $a$.
}
\textcolor{black}{
For the infinitesimal gauge transformation with parameters $\omega^{a}$, $\Omega(s)$ can be expressed as $\Omega(s) = e^{i\omega^{a}T^{a}} \simeq 1 + i\omega^{a}T^{a} + O(\omega^{2})$.
Thus, in the continuous limit, Eq.\eqref{eq:gt_lat_infinitesimal} is written as
\begin{eqnarray}
    \label{eq:gt_lat_infinitesimal2}
    \delta R^{\mathrm{lat}}_{\mathrm{DR}}
    & = &
    - ag
    \mathrm{Re Tr}
    \left[
        \omega^{a} T^{a}
        \sum_{\perp = x,y}
        \left(
            A_{\perp}(s) - A_{\perp}(s-a_{\perp})
        \right)
    \right] \nonumber \\
    & = &
    - a^{2} g
    \mathrm{Re Tr}
    \left[
        \omega^{a} T^{a}
        \sum_{\perp = x,y}
        \left(
            \partial^{\mathrm{B}}_{\perp} A_{\perp}(s)
        \right)
    \right]
\end{eqnarray}
up to $O(a^{2})$.
From the first to second lines, we use a backward derivative $\partial^{\mathrm{B}}_{\perp}$.
The extremum condition $\delta R^{\mathrm{lat}}_{\mathrm{DR}} = 0$ for any $\omega^{a}$ leads to
\begin{equation}
    \label{eq:DRG_lat_local}
    \sum_{\perp = x,y}
    \partial^{\mathrm{B}}_{\perp} A_{\perp}(s)
    =
    0.
\end{equation}
}
Thus, the local condition \eqref{eq:DRG_local} is derived in the continuous limit.

Now, we consider $tz$-projection in lattice QCD.
In continuous spacetime, we define the $tz$-projection as the replacement $A_{x,y}(s) \to 0$.  
In lattice QCD, the $tz$-projection is defined by a simple replacement:
\begin{equation}
    \label{eq:tz_proj_lat}
    U_{x,y}(s) \to 1. 
\end{equation}

The $tz$-projection changes the action \eqref{eq:plaquette} into
\begin{eqnarray}
    \label{eq:DRG_lat_act_tzp}
    && \hspace{-20pt}
    S^{\mathrm{lat}}_{tz\mathrm{-DR}} 
    = 
    \beta \sum_{s}
    \Biggl[
    \left\{
    1 - \frac{1}{N_{c}} \mathrm{ReTr}P_{tz}(s)
    \right\} 
    \nonumber \\
    && \hspace{-10pt}
    + \hspace{-3pt}
    \left.
    \sum_{\mu = t, z} \hspace{-3pt}
    \left\{
        1 - \frac{1}{N_{c}} \hspace{-2pt}
        \sum_{\perp = x,y} \hspace{-5pt}
        \mathrm{ReTr} \hspace{-2pt}
        \left[
            U_{\mu}(s) U^{\dagger}_{\mu}(s+a_{\perp})
        \right]
    \right\}
    \right]
\end{eqnarray}
at the tree-level.
As well as the action \eqref{eq:DRG_act_tzp}, the first term is 
the 2D lattice YM action on the $t$-$z$ plane, and the second term is interpreted as interaction between neighbors in the $x$ and $y$ directions:
the interaction is written as the product of neighboring link-variables, $U_{\mu}(s) U^{\dagger}_{\mu}(s+a_{\perp})$, for $\mu=t, z$.
Then, through this interaction, 2D YM systems seem to be correlated in the $x$ and $y$ directions.

In the next section, we perform  
DR gauge fixing and $tz$-projection in lattice QCD. 
In the practical calculation, 
the $tz$-projection is achieved 
by removal of $A_x$ and $A_y$, i.e., 
replacement of $U_x$ and $U_y$ by unity, 
for the gauge configuration 
generated in lattice QCD in the DR gauge, 
in a similar manner to Abelian projection 
in the MA gauge 
\cite{KRONFELD_MAG2_1987, SUZUKI_YOTSUYANAGI_AD, 
Brandstaeter_MonopoleCond_1991, 
Miyamura_MonopoleChralCond_1995, Woloshyn_MonopoleChralCond_1995, 
Chernodub_Polikarpov_1997, 
Ichie_Suganuma_1999, Amemiya_Suganuma_1999, Gongyo_Iritani_Suganuma_2012, 
SAKUMICHI_SUGANUMA_AD, 
Sakumichi_Suganuma_AD_2015, Ohata_Suganuma_MonopoleChralCond_2021} or 
center projection in the maximal center gauge 
\cite{Center_Vortex_1997}.

\subsection{Comparison of DR gauge with MA gauge}
\label{subsec:comparison_DR_MA}
Before proceeding the lattice QCD calculation,
we compare the DR gauge with the MA gauge in this subsection.
Using the Cartan subalgebra $H^{a}$ and the raising and lowering operators $E_{\pm \alpha}$ of $\mathfrak{su}(N_{c})$, the gluon field $A_{\mu}(s) \in \mathfrak{su}(N_{c})$ can be expressed as
\begin{equation}
    \label{eq:cartan}
    A_{\mu}(s)
    =
    A^{i}_{\mu}(s) H^{i}
    +
    A^{\alpha}_{\mu}(s) E_{-\alpha},
\end{equation}
where $A^{i}_{\mu}(s)$ denotes the diagonal component and $A^{\alpha}_{\mu}(s)$ the off-diagonal component of the gluon field.
The MA gauge is defined so as to minimize
\begin{equation}
    \label{eq:MAG}
    R_{\mathrm{MA}}
    \equiv
    \int d^{4} s
    \sum_{\alpha}
    \left[
    A^{\alpha}_{\mu}(s) A^{-\alpha}_{\mu}(s)
    \right]
\end{equation}
by gauge transformation in Euclidean spacetime.
Comparing $R_{\mathrm{MA}}$ with $R_{\mathrm{DR}}$ in Eq.\eqref{eq:DRG}, they take similar form.
The summation is taken for the internal color index $\alpha$ in the MA gauge, while the sum for the external spacetime index $\perp$ in the DR gauge.
Accordingly, the residual gauge symmetry becomes $\mathrm{U}(1)^{N_{c}-1}$ in the MA gauge, while the symmetry for the gauge function $\Omega(t,z)$.
From the definition with Eq.\eqref{eq:MAG}, the amplitude of off-diagonal components $A^{\alpha}_{\mu}(s)$ is strongly suppressed in the MA gauge, which is demonstrated in lattice QCD \cite{Ichie_Suganuma_1999, 
Sakumichi_Suganuma_AD_2015}.

From lattice QCD calculations, it is also found that, in the MA gauge, only the diagonal components $A^{i}_{\mu}(s)$ play dominant role to the low-energy phenomena such as quark confinement \cite{SUZUKI_YOTSUYANAGI_AD, 
SAKUMICHI_SUGANUMA_AD, 
Sakumichi_Suganuma_AD_2015} and spontaneous chiral symmetry breaking \cite{Miyamura_MonopoleChralCond_1995, Woloshyn_MonopoleChralCond_1995, Ohata_Suganuma_MonopoleChralCond_2021}, which is called Abelian dominance \cite{EZAWA_Iwazaki_AD_1982}.
On the other hand, the off-diagonal components $A^{\alpha}_{\mu}(s)$ become massive in the MA gauge, and their mass is estimated as $M_{\mathrm{off}} \sim 1 \; \mathrm{GeV}$ \cite{Amemiya_Suganuma_1999, Gongyo_Iritani_Suganuma_2012}.
Thus, long-range propagation of off-diagonal gluons is strongly suppressed and Abelian dominance is realized in the MA gauge.

From the similarity, we might expect that two gauge components $A_{t}(s)$ and $A_{z}(s)$ are dominant in the low-energy region, and the other components $A_{x}(s)$ and $A_{y}(s)$ do not make a major contribution to quark confinement in the DR gauge.

\section{NUMERICAL CALCULATION IN LATTICE QCD}
\label{sec:numerical}
To investigate effective dimensional reduction in the 4D DR-gauged YM theory, we perform $\mathrm{SU}(3)$ lattice QCD Monte Carlo calculations at the quenched level. 
We use the standard plaquette action with $\beta = 6.0$ on a 4D lattice of size $N_{s}=24^{4}$. 
The lattice spacing is $a \simeq 0.10 \; \mathrm{fm}$, i.e., $a^{-1} \simeq 2.0 \; \mathrm{GeV}$ at $\beta = 6.0$,
which is determined from the string tension \cite{Takahashi_Suganuma_detailedQQpot_2002}.
Using the pseudo-heat bath algorithm, we generate 800 gauge configurations, 
which are picked up every 1000 sweeps after 20,000 sweeps for thermalization.

After generating the gauge configurations, we perform DR gauge fixing.
For each gauge configuration, we numerically maximize  $R^{\mathrm{lat}}_{\mathrm{DR}}$ in Eq.\eqref{eq:DRG_lat}  
using an iterative maximization algorithm, 
similarly in Landau or MA gauge fixing \cite{Sakumichi_Suganuma_AD_2015, Ohata_Suganuma_MonopoleChralCond_2021}. 
For rapid convergence, we use the over-relaxation (OR) method 
with the OR parameter $1.6$. 
When $R^{\mathrm{lat}}_{\mathrm{DR}}$ is maximized, 
\begin{equation}
    \label{eq:gfix_violation}
    \Delta_{\mathrm{DR}}(s)
    \equiv 
    \sum_{\perp = x, y}
    \partial^{\mathrm{B}}_{\perp} \mathcal{A}_{\perp}(s)
\end{equation}
has to be zero.
Here, the lattice gluon field is defined as 
\begin{equation}
    \label{eq:lat_gauge_field}
    \mathcal{A}_{\perp}(s)
    \equiv 
    \frac{1}{2iag} 
    \left.
    \left[
        U_{\perp}(s) - U^{\dagger}_{\perp}(s) 
    \right]
    \right|_{\mathrm{traceless}} 
    \in
    \mathfrak{su}(N_{c}),
\end{equation}
where ``traceless'' means the subtraction of its trace part. 
As a numerical convergence criterion, we impose that the maximization of $R^{\mathrm{lat}}_{\mathrm{DR}}$ stops when 
$\epsilon_{\mathrm{DR}} < 4.0 \times 10^{-12}$ is satisfied
for 
\begin{equation}
    \label{eq:gfix_violation_norm}
    \epsilon_{\mathrm{DR}}
    \equiv 
    \frac{1}{N_{c} N_{s}}
    \sum_{s}
    \mathrm{Tr}
    \left[
        \Delta_{\mathrm{DR}}(s)
        \Delta^{\dagger}_{\mathrm{DR}}(s)   
    \right].
\end{equation}

Finally in this subsection, 
we try a simple numerical check 
\cite{SAKUMICHI_SUGANUMA_AD}
on the Gribov ambiguity \cite{Gribov_1978} 
in the DR gauge fixing.
The DR gauge fixing is 
achieved by maximizing the global value of 
$R_{\rm DR}^{\rm lat}[U_\perp]$ 
in Eq.(\ref{eq:RDRlat}), which leads to 
the local gauge fixing condition (\ref{eq:DRG_lat_local}). 
The Gribov ambiguity is expressed as 
taking a ``bad local maximum" 
where $R_{\rm DR}^{\rm lat}$ is unacceptably small 
or accidentally taking a ``local minimum" of $R_{\rm DR}$, which also satisfies the local gauge fixing condition (\ref{eq:DRG_lat_local}). 
Then, we check the value of $R_{\rm DR}^{\rm lat}$ for each gauge configuration in lattice QCD. 
For, when a bad maximum or a local minimum is taken in some gauge configuration, the value of $R_{\rm DR}$ becomes too small than the average.
For 800 used gauge configurations,
we calculate the ensemble averaged value $\langle R_{\rm DR}^{\rm lat}\rangle$ and
its standard deviation $\Delta R_{\rm DR}^{\rm lat}$ in the DR gauge.
For the normalized quantity $R_{\rm DR}^{\rm lat}/(2 L^3 L_t N_c)$, 
which is unity when $U_\perp \equiv 1$, we find 
\begin{eqnarray}
\label{eq:Ave_R_DR}
\frac{1}{2 L^3 L_t N_c}\langle R_{\rm DR}^{\rm lat}\rangle &\simeq& 0.921, \\
\label{eq:Std_R_DR}
\frac{1}{2 L^3 L_t N_c} \Delta R_{\rm DR}^{\rm lat} &\simeq& 0.972\times 10^{-4}.
\end{eqnarray}
In comparison with the ensemble average $\langle R_{\rm DR}^{\rm lat}\rangle$, 
the standard deviation $\Delta R_{\rm DR}^{\rm lat}$ is fairly small.
In other words, the maximized value of $R_{\rm DR}^{\rm lat}$ 
is almost the same for all the gauge configurations. 
This seems to indicate that the present method succeeds to avoid bad local extreme where $R_{\rm DR}^{\rm lat}$ is relatively small.
Then, we expect that the Gribov copy effect is not so significant in our calculation.  

\subsection{Properties of link-variables in DR gauge}
\label{subsec:link_distance}
As the local property of link-variables in the DR gauge, we define and compute a distance between a link-variable $U_{\mu}$ and a unit matrix $I$.
Denoting the distance $d(U_{\mu},I)$,
the squared distance is defined as 
\begin{eqnarray}
    \label{eq:dist_link}
    d(U_{\mu}, I)^{2} 
    & = &
    \frac{1}{2N_{c}} \mathrm{Tr}\left[ 
    (U_{\mu}-I)^{\dagger}
    (U_{\mu}-I)
    \right] \nonumber \\
    & = &
    \frac{1}{2N_{c}}
    \sum^{N_{c}}_{i,j=1}
    \left| (U_{\mu}-I)_{ij} \right|^{2} ,
\end{eqnarray}
which is proportional to the Frobenius norm of the matrix $U_{\mu}-I$.
Since $U_{\mu}$ is an element of $\mathrm{SU}(N_{c})$, 
\begin{eqnarray}
    \label{eq:dist_link2}
    d(U_{\mu}, I)^{2} 
    & = &
    \frac{1}{2N_{c}} \mathrm{Tr}
    \left(
        2I - U_{\mu} - U_{\mu}^{\dagger}
    \right) \nonumber \\
    & = &
    1 
    -
    \frac{1}{N_{c}}
    \mathrm{ReTr} \; U_{\mu} \; .
\end{eqnarray}

For even $N_{c}$, $-I$ is an element of $\mathrm{SU}(N_{c})$, and the maximum value of $d(U_{\mu}, I)^{2}$ is realized for $U_{\mu} = - I$: $d(U_{\mu}=-I, I)^{2} = 2$.
On the other hand, for odd $N_{c}$, $U_{\mu}=-I$ does not belong to $\mathrm{SU}(N_{c})$.
In this case, the maximum value of $d(U_{\mu}, I)^{2}$ is realized 
when $U_{\mu}$ is the closest element to $-I$ 
among the center of  $\mathbb{Z}^{N_{c}} \in {\rm SU}(N_c)$. 
In fact, the range of $d(U_{\mu}, I)^{2}$ is found to be
\begin{equation}
    \label{eq:dist_range}
    \begin{cases}
        0 \le d(U_{\mu}, I)^{2} \le 2 & (N_{c} \textrm{ is even}), \\
        0 \le d(U_{\mu}, I)^{2} \le 
        1 - \cos \frac{\pi ( N_{c} - 1 )}{N_{c}}
        & (N_{c} \textrm{ is odd}) .
    \end{cases}
\end{equation}

Table \ref{tab:distance} shows the vacuum expected value of $d(U_{\mu}, I)^2$ calculated in SU(3) lattice QCD .
In the case of no gauge fixing, we find $\langle d(U_{\mu}, I)^{2} \rangle = 1$, which can be analytically shown with the integration on the $\mathrm{SU}(N_{c})$ group manifold.
As for the result of $\langle d(U_{t,z}, I)^{2} \rangle_{\mathrm{DR}}$ in the second line, similar argument for the residual gauge symmetry with $\Omega(t,z)$ can be applied, and the distance is calculated as unity.
(In this paper, we use 
$\langle \cdots \rangle_{\mathrm{DR}}$ for a vacuum expectation value in the DR gauge.) 

On the other hand, the vacuum expectation value $\langle d(U_{\perp}, I)^{2} \rangle_{\mathrm{DR}}$ equals to $0.076$ in the DR gauge, which is about one order smaller than the value of no gauge fixing.
Therefore, the amplitudes of two components $A_{x}(s)$ and $A_{y}(s)$ are strongly suppressed by the DR gauge fixing.

\begin{center}
\begin{table}[h]
    \centering
    \begin{tabular}{cc} 
    \hline
        gauge & $ \langle d(U_\mu, I)^{2} \rangle $  \\ 
    \hline
        No fixing & 1.000 \\
        DR $(\mu = t,z)$ & 1.000 \\
        DR $(\perp=x,y)$ & 0.076 \\
    \hline
    \end{tabular}
    \caption{
    Vacuum expectation values of $d(U_{\mu}, I)^{2}$.
    The first line is for no gauge fixing, the second line for $U_{t,z}$ in the DR gauge, and the third line for $U_{x,y}$ in the DR gauge.
    }
    \label{tab:distance}
\end{table}
\end{center}

Note that, the value of 
$\langle d(U_{\perp}, I)^{2} \rangle_{\rm DR}$  
seems to be consistent with Eq.(\ref{eq:Ave_R_DR}).
From Eq.\eqref{eq:dist_link2}, $\langle d(U_{\perp}, I)^{2} \rangle_{\mathrm{DR}}$ can be written as
\begin{equation}
    \label{eq:consistent_Gribov_d}
    \langle d(U_{\perp}, I)^{2} \rangle_{\mathrm{DR}}
    =
    1
    - \frac{1}{N_{c}}\langle \mathrm{ReTr} \; U_{\perp} \rangle_{\mathrm{DR}}
    \simeq 0.076.
\end{equation}
Then, the summation of $\langle R_{\rm DR}^{\rm lat}\rangle$ and $\langle d(U_{\perp}, I)^{2} \rangle_{\mathrm{DR}}$ is to be unity 
under appropriate normalization,
\begin{equation}
    \label{eq:consistent_unity}
    \frac{1}{2 L^3 L_t N_c}
    \left[
    \langle R_{\rm DR}^{\rm lat}\rangle
    +
    \sum_{s} \sum_{\perp = x,y} \langle d(U_{\perp}, I)^{2} \rangle_{\mathrm{DR}}
    \right]
    =
    1 \;.
\end{equation}
From Eqs.(\ref{eq:Ave_R_DR}) and (\ref{eq:consistent_Gribov_d}), 
the lattice result of the sum is found to be almost unity,
\begin{equation}
    \label{eq:consistent_numerical_unity}
    0.921 + 0.076 \simeq 0.997 \simeq 1,
\end{equation}
which indicates consistency of the lattice calculation.

\subsection{Wilson loop and interquark potential after $tz$-projection in DR-gauged YM theory}
\label{subsec:WilsonLoop_DR_proj.ed}

As shown in the previous section, $A_{x}(s)$ and $A_{y}(s)$ 
are strongly suppressed in the DR gauge.

From the similarity between DR gauge and MA gauge as discussed in Sec.\ref{subsec:comparison_DR_MA},
we might expect that two gluon components $A_{t}(s)$ and $A_{z}(s)$ play a dominant role in low-energy phenomena such as quark confinement.
To investigate this, we apply the $tz$-projection to the Wilson loop and extract the interquark potential.

\begin{figure}[h]
    \centering
    \includegraphics[width=6.5cm]{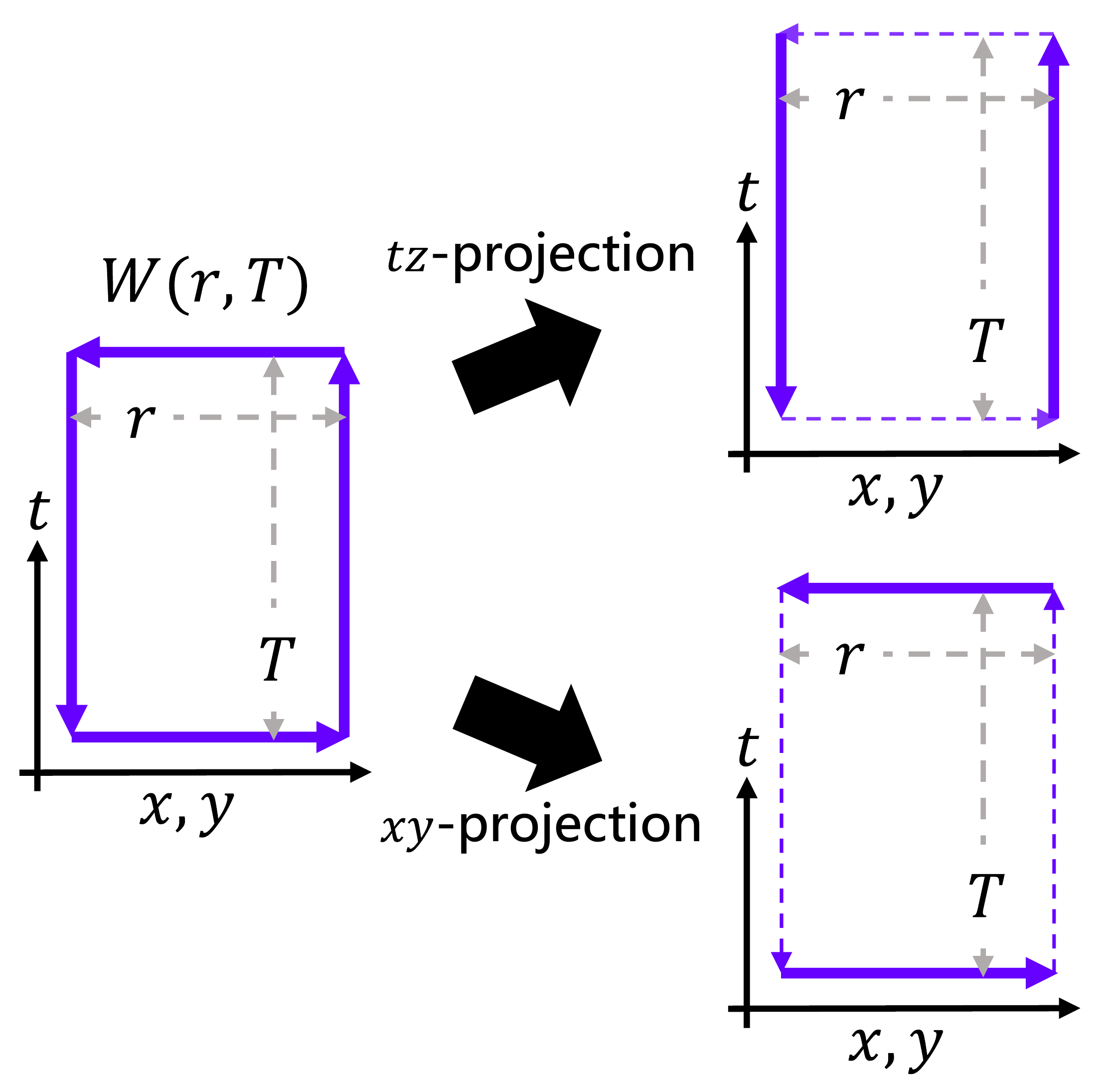}
    \caption{
    Schematic figure of the $tz$-projection and $xy$-projection to the Wilson loop on the $t$-$\perp$ plane, respectively.
    }
    \label{fig:projected_wlp}
\end{figure}

As the opposite of the $tz$-projection, we define ``$xy$-projection" as replacement
\begin{equation}
    \label{eq:xy_proj}
    A_{t,z} \to 0 \:
    \Leftrightarrow  \:
    U_{t,z} \to 1.
\end{equation}
We apply the $tz$-projection and the $xy$-projection to the Wilson loop as shown in Fig.\ref{fig:projected_wlp}.

Note that the $tz$-projection does not change the Wilson loop on the $t$-$z$ plane 
because it does not contain $U_{\perp} \; (\perp = x,y)$ explicitly. 
Then, we consider the Wilson loop on the $t$-$\perp \; (\perp = x,y)$ plane.
(Of course, the expectation value of the Wilson loop on the $t$-$z$ plane 
is the same as that on the $t$-$\perp$ plane, 
although the description looks different in the DR gauge.)

Before showing numerical results, we mention about the gauge transformation property of these projected Wilson loops in terms of the residual gauge transformation with $\Omega(t,z)$.

\begin{figure}[h]
    \centering
    \includegraphics[width=7.0cm]{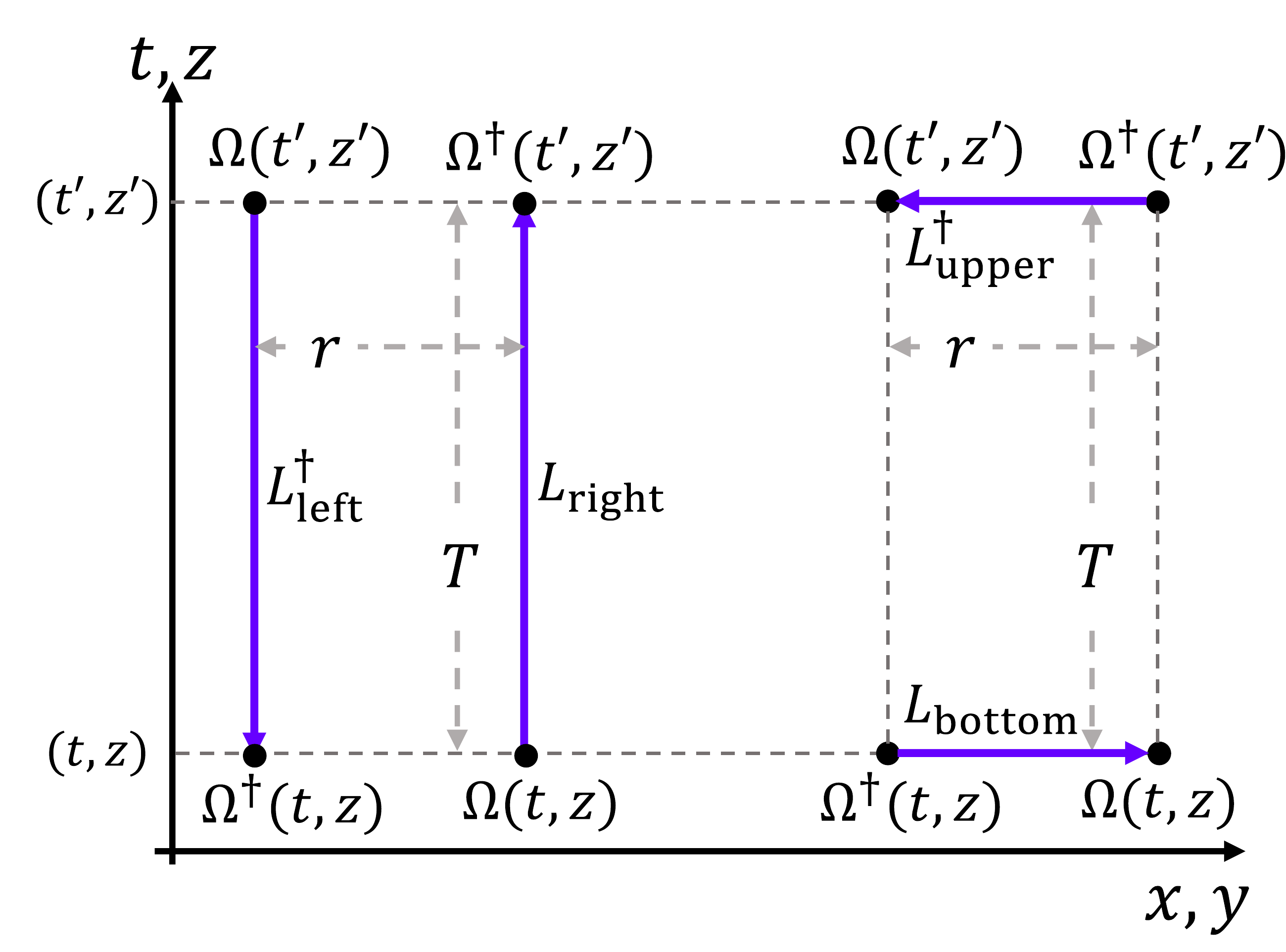}
    \caption{
    The $tz$-projected Wilson loop (left) and the $xy$-projected Wilson loop (right) on the $t$-$\perp$ plane.
    We note $t' = t + T$ and $z' = z$.
    }
    \label{fig:projected_wilsonloop_explain}
\end{figure}

Figure \ref{fig:projected_wilsonloop_explain} shows the $tz$- and $xy$-projected Wilson loops on the $t$-$\perp$ plane.
First, we consider the $tz$-projected Wilson loop $W^{tz}(r,T)$, the left one in Fig.\ref{fig:projected_wilsonloop_explain}.
This Wilson loop is decomposed into two Wilson lines  $L^{\dagger}_{\mathrm{left}}$ and $L_{\mathrm{right}}$, 
\begin{equation}
    \label{eq:tz_wilsonloop}
    W^{tz}(r,T)
    =
    \mathrm{Tr}
    \left[
        L^{\dagger}_{\mathrm{left}} L_{\mathrm{right}}
    \right],
\end{equation}
as shown in Fig.\ref{fig:projected_wilsonloop_explain}.
Under the residual gauge transformation with $\Omega(t,z)$,
$L^{\dagger}_{\mathrm{left}}$ and $L_{\mathrm{right}}$ transform as
\begin{eqnarray}
    \label{eq:res_gt_left_wilsonline}
    L^{\dagger}_{\mathrm{left}}
    & \to &
    \Omega(t',z') L^{\dagger}_{\mathrm{left}} \Omega^{\dagger}(t,z) , \\
    \label{eq:res_gt_right_wilsonline}
    L_{\mathrm{right}}
    & \to &
    \Omega(t,z) L_{\mathrm{right}} \Omega^{\dagger}(t',z') ,
\end{eqnarray}
and the $tz$-projected Wilson loop transforms as 
\begin{eqnarray}
    \label{eq:res_gt_tz_wilsonloop}
    W^{tz}(r,T)
    & \to &
    \mathrm{Tr}
    \left[
        \Omega(t',z') L^{\dagger}_{\mathrm{left}} \Omega^{\dagger}(t,z)
        \Omega(t,z) L_{\mathrm{right}} \Omega^{\dagger}(t',z')
    \right] \nonumber \\
    & = &
    \mathrm{Tr}
    \left[
        L^{\dagger}_{\mathrm{left}} 
        L_{\mathrm{right}}
    \right]
    =
    W^{tz}(r,T) .
\end{eqnarray}
Thus, the $tz$-projected Wilson loop is invariant under the residual gauge transformation.

Next, we consider the $xy$-projected Wilson loop $W^{xy}(r,T)$, the right one in Fig.\ref{fig:projected_wilsonloop_explain}.
This Wilson loop is also decomposed into two Wilson lines $L^{\dagger}_{\mathrm{upper}}$ and $L_{\mathrm{bottom}}$,
\begin{equation}
    \label{eq:xy_wilsonloop}
    W^{xy}(r,T)
    =
    \mathrm{Tr}
    \left[
        L^{\dagger}_{\mathrm{upper}} L_{\mathrm{bottom}}
    \right].
\end{equation}
Under the residual gauge transformation with $\Omega(t,z)$, these two Wilson lines transform as
\begin{eqnarray}
    \label{eq:res_gt_top_wilsonline}
    L^{\dagger}_{\mathrm{upper}}
    & \to &
    \Omega(t',z') L^{\dagger}_{\mathrm{upper}} \Omega^{\dagger}(t',z'), \\
    \label{eq:res_gt_bottom_wilsonline}
    L_{\mathrm{bottom}}
    & \to &
    \Omega(t,z) L_{\mathrm{bottom}} \Omega^{\dagger}(t,z) .
\end{eqnarray}
Therefore, the $xy$-projected Wilson loop transforms as
\begin{eqnarray}
    \label{eq:res_gt_xy_wilsonloop}
    W^{xy}(r,T)
    & \to &
    \mathrm{Tr}
    \left[
        \Omega(t',z') L^{\dagger}_{\mathrm{upper}} \Omega^{\dagger}(t',z')
    \right. \nonumber \\
    && \hspace{20pt}
    \left.
        \Omega(t,z) L_{\mathrm{bottom}} \Omega^{\dagger}(t,z)
    \right] \nonumber \\
    & = &
    \mathrm{Tr}
    \left[
        \tilde{\Omega} L^{\dagger}_{\mathrm{upper}} 
        \tilde{\Omega}^{\dagger} L_{\mathrm{bottom}}
    \right] ,
\end{eqnarray}
where $\tilde{\Omega} \equiv \Omega^{\dagger}(t,z) \Omega(t',z')$.
After some consideration in Appendix \ref{sec:xy_proj_Wlp}, we find that the vacuum expectation value of $\langle W^{xy}(r,T) \rangle_{\mathrm{DR}}$ is calculated as
\begin{equation}
    \label{eq:res_gt_xy_wilsonloop3}
    \langle 
        W^{xy}(r,T)
    \rangle_{\mathrm{DR}}
    =
    \frac{1}{N_{c}}
    \left\langle 
        \mathrm{Tr} \; L^{\dagger}_{\mathrm{upper}}
        \mathrm{Tr} \; L_{\mathrm{bottom}}
    \right\rangle_{\mathrm{DR}}.
\end{equation}

\subsubsection{Interquark potential from $tz$-projected Wilson loop}
\label{subsubsubsec:tz_proj.ed_Wilson}
Now, we investigate the effect from the $tz$-projection of $U_{x,y}(s) \to 1$ for the interquark potential.
To this end, we calculate the static interquark potential from the $tz$-projected Wilson loop 
$\langle W^{tz}(r,T) \rangle_{\mathrm{DR}}$ 
on the $t$-$\perp$ plane in the DR gauge.
In fact, starting from ordinary lattice QCD Monte Carlo sampling, 
we compute the Wilson loop 
using the $tz$-projected gauge  configurations in the DR gauge. 
Similar to the ordinary static potential, we define the $tz$-projected static potential $V^{tz}(r)$ as
\begin{equation}
    \label{eq:potential_tz_proj}
    \langle W^{tz}(r,T) \rangle_{\mathrm{DR}}
    =
    A e^{-V^{tz}(r)T}
\end{equation}
for large $T$.
Then, the $tz$-projected potential $V^{tz}(r)$ is extracted from $\langle W^{tz}(r,T) \rangle_{\mathrm{DR}}$ as
\begin{equation}
    \label{eq:potential_tz_proj2}
    V^{tz}(r)
    =
    - \frac{1}{T}
    \ln \langle W^{tz}(r,T) \rangle_{\mathrm{DR}}.
\end{equation}
For the accuracy and efficiency of numerical calculations, we have used the gauge-covariant smearing method with Refs.\cite{APE_smearing_1987, Takahashi_Suganuma_detailedQQpot_2002}.

\begin{figure}[h]
    \centering
    \includegraphics[width=9.0cm]{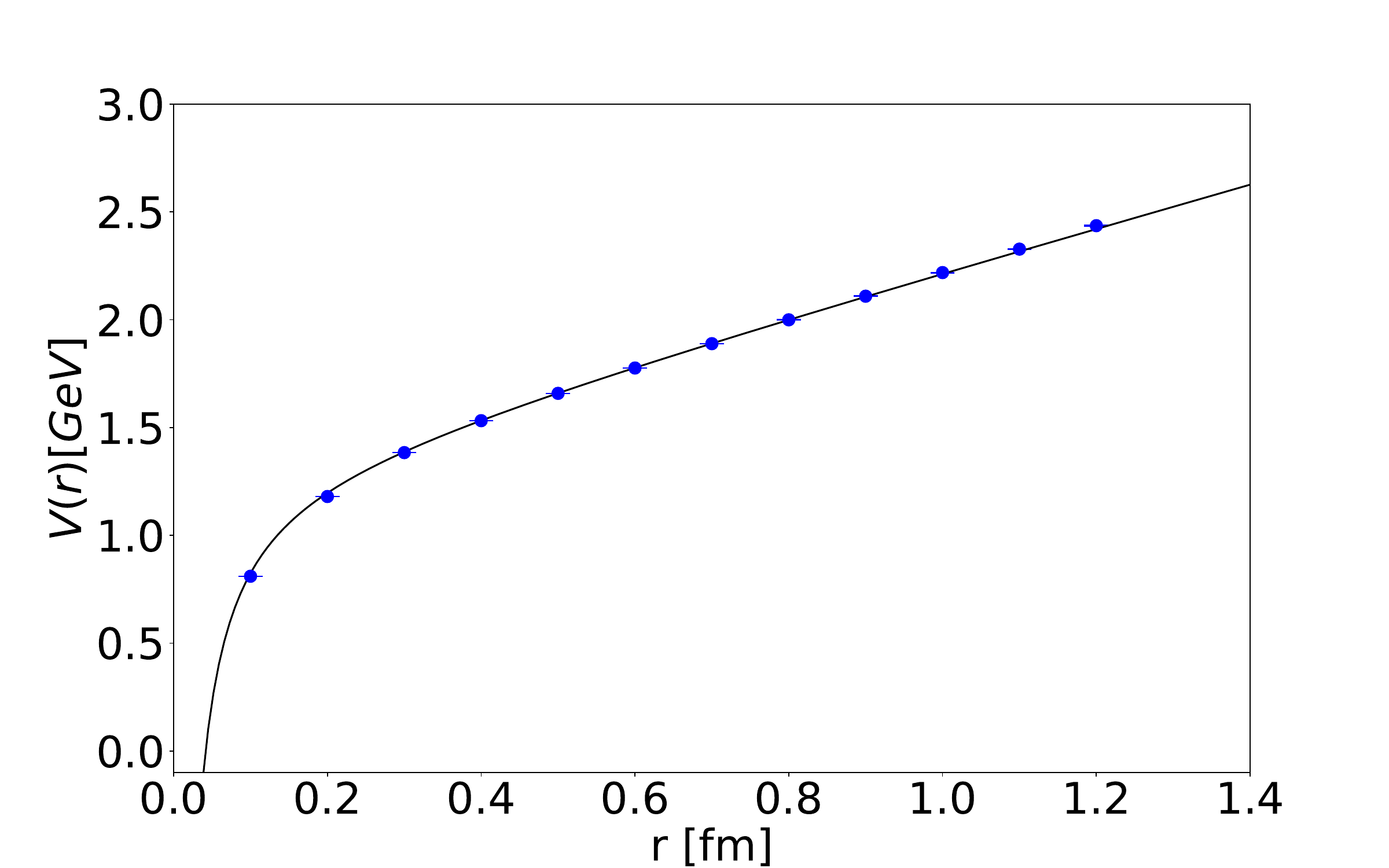}
    \caption{
    The interquark potential $V^{tz}(r)$ from the $tz$-projected Wilson loops $\langle W^{tz}(r,T) \rangle_{\mathrm{DR}}$
    on the $t$-$\perp$ plane.
    The horizontal axis $r$ denotes the interquark distance, and the vertical axis the potential energy.
    The dots denote the $tz$-projected potential $V^{tz}(r)$ in lattice QCD at $\beta = 6.0$, and the solid line the best fit Cornell potential in lattice QCD \cite{Takahashi_Suganuma_detailedQQpot_2002}. 
    }
    \label{fig:wilson_tzp}
\end{figure}

The lattice QCD result is shown in Fig.\ref{fig:wilson_tzp}.
The horizontal axis $r$ denotes the interquark distance, and the vertical axis the potential energy.
The dots denote the $tz$-projected potential $V^{tz}(r)$ 
calculated from $\langle W^{tz}(r,T) \rangle_{\rm DR}$
on the $t$-$\perp$ plane, and the solid line the standard interquark potential calculated in $\mathrm{SU}(3)$ lattice QCD \cite{Takahashi_Suganuma_detailedQQpot_2002} of which the functional form is found to be the Cornell potential \cite{Cornell}.
The $tz$-projected potential $V^{tz}(r)$ is good agreement with the Cornell potential.
This means that the interquark potential is reproduced 
with two gluon components $A_{t}(s)$ and $A_{z}(s)$ 
in the DR gauge.

The dominant role of $A_{t}(s)$ and $A_{z}(s)$  for the static potential seems to be natural because only the temporal gauge component is relevant for it \cite{Lüscher_Weisz_2001}. 
However, this result is practically nontrivial at least for the terminated Wilson-line correlator in lattice QCD, 
since the static potential cannot be reproduced only with the temporal gluon, e.g., in the Landau gauge 
 \cite{Iritani:generalized_landau}.

\subsubsection{$xy$-projected Wilson loop}
\label{subsubsec:xy_proj.ed_Wilson}
In the previous section, we have found that the interquark potential is reproduced with two gauge components $A_{t}(s)$ and $A_{z}(s)$ in the DR gauge.
Then, we investigate a contribution from $A_{x}(s)$ and $A_{y}(s)$ to quark confinement.
We calculate the $xy$-projected Wilson loop $\langle W^{xy}(r, T) \rangle_{\mathrm{DR}}$.

\begin{figure}[h]
    \centering
    \includegraphics[width=9.0cm]{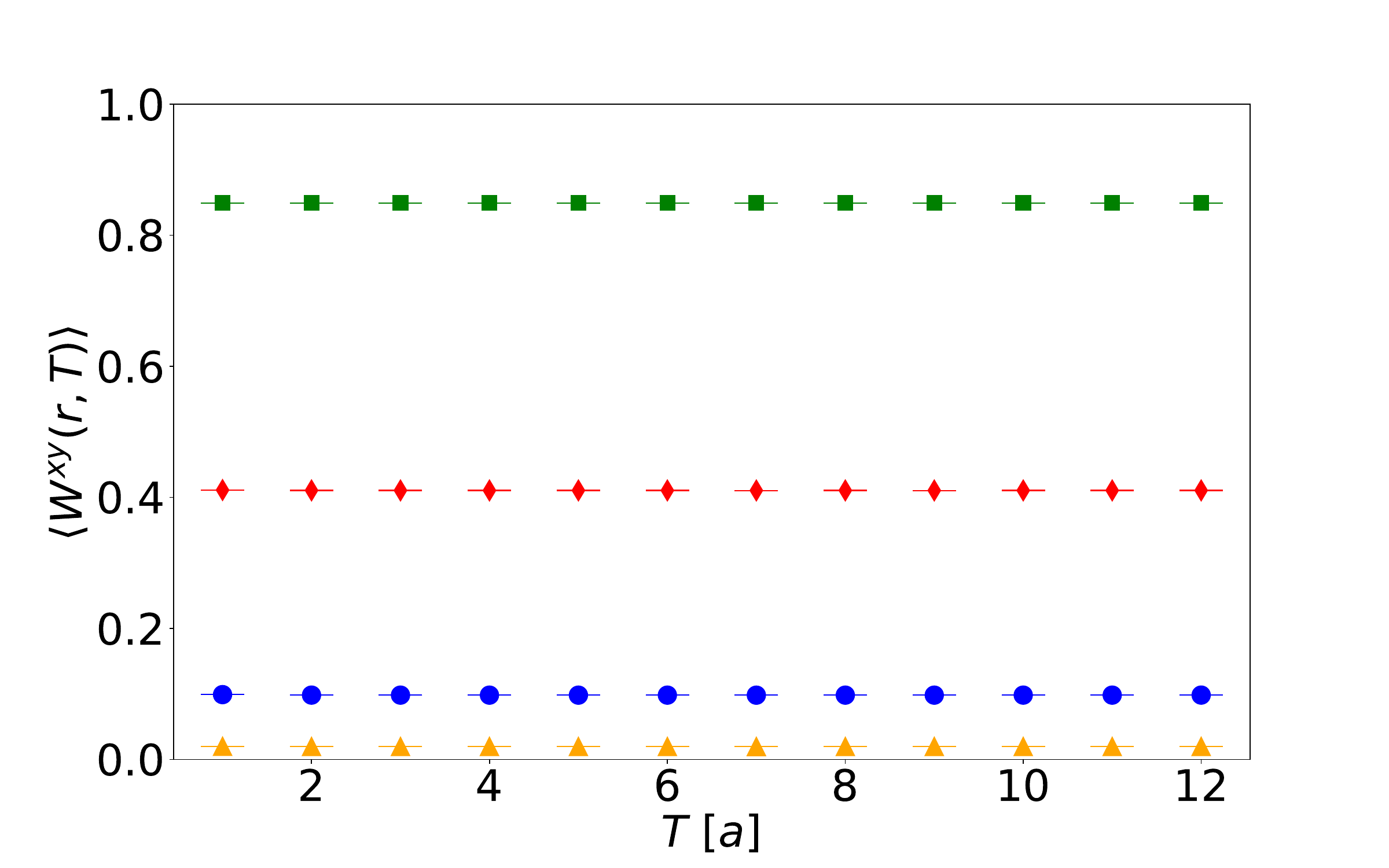}
    \caption{
    The lattice QCD result of the $xy$-projected Wilson loop $\langle W^{xy}(r, T) \rangle_{\mathrm{DR}}$ in the DR gauge, 
    plotted against the temporal length $T$, 
    for several values of the spatial length $r$, i.e.,  
    $r=1$ (green square), $r=3$ (red diamond), $r=6$ (orange circle) and $r=9$ (orange triangle).
    }
    \label{fig:wilson_xyp}
\end{figure}

Figure \ref{fig:wilson_xyp} shows the lattice QCD result at $\beta = 6.0$. 
The vertical axis denotes the $xy$-projected Wilson loop $\langle W^{xy}(r, T)$, and the horizontal axis the temporal length $T$.
Denoting the spatial length $r$ of the Wilson loop, we plot $\langle W^{xy}(r, T) \rangle_{\mathrm{DR}}$ for $r=1$ (green square), $r=3$ (red diamond), $r=6$ (orange circle) and $r=9$ (orange triangle) as typical distances.

The $xy$-projected Wilson loop is independent of $T$:
\begin{equation}
    \label{eq:const_wilson_xy}
    \langle
        W^{xy}(r, T)
    \rangle_{\mathrm{DR}}
    =
    \phi(r).
\end{equation}
We also define the $xy$-projected static potential $V^{xy}(r)$ as
\begin{equation}
    \label{eq:const_wilson_xy2}
    \langle
        W^{xy}(r, T)
    \rangle_{\mathrm{DR}}
    =
    e^{-V^{xy}(r) T}
\end{equation}
for large $T$.
Then, the $xy$-projected potential $V^{xy}(r)$ is calculated from $\langle W^{xy}(r, T) \rangle_{\mathrm{DR}}$ as
\begin{equation}
    V^{xy}(r) 
    =
    - \lim_{T \to \infty}
    \frac{1}{T}
    \ln 
    \langle
        W^{xy}(r, T)
    \rangle_{\mathrm{DR}}.
\end{equation}
Because of Eq.\eqref{eq:const_wilson_xy}, we find
\begin{equation}
    \label{eq:zero_pot_xy}
    V^{xy}(r) 
    =
    0 .
\end{equation}
This result suggest that $A_{x}(s)$ and $A_{y}(s)$ do not make major contribution to quark confinement in the DR gauge.

As a caution, since the link-variables are not commutative, the Wilson loop cannot be simply factorized into $W^{tz}(r, T)$ and $W^{xy}(r, T)$:
\begin{equation}
    \label{eq:fact_Wlp}
    W(r, T)
    \neq
    W^{tz}(r, T)
    W^{xy}(r, T).
\end{equation}
Therefore, $A_{x}(s)$ and $A_{y}(s)$ might give some contribution to the whole Wilson loop, although their contribution would be small for quark confinement. 

\subsection{Spatial correlation and mass of $A_{x}(s)$ and $A_{y}(s)$ in DR gauge}
\label{subsec:transverse_mass}
In Sec.\ref{subsec:WilsonLoop_DR_proj.ed}, it has been shown that two gauge component $A_{t}(s)$ and $A_{z}(s)$ play a dominant role in quark confinement, and the other components $A_{x}(s)$ and $A_{y}(s)$ does not contribute.
Here, we consider the reason why $A_{x}(s)$ and $A_{y}(s)$ are inactive in the infrared region.

In the MA gauge, the large mass of the off-diagonal components $A^{\alpha}_{\mu}(s)$ is considered to realize infrared inactivity \cite{Amemiya_Suganuma_1999, Gongyo_Iritani_Suganuma_2012}.
Here, we calculate the spatial correlation of two $\perp$-directed link-variables and estimate the mass of $A_{x}(s)$ and $A_{y}(s)$ in the DR gauge.

The gluon mass can be estimated from the gluon propagator $G(r)$ which is defined as a two-point function of gluon fields,
\begin{equation}
    \label{eq:gluon_propagator}
    G(r)
    \equiv
    \langle
        A^{a}_{\perp}(0) A^{a}_{\perp}(ra_{\perp})
    \rangle_{\mathrm{DR}},
\end{equation}
where we have used the translational symmetry in the $\perp$ directions. 
As shown in Sec.\ref{subsec:link_distance}, the amplitudes of $A_{x}(s)$ and $A_{y}(s)$ are strongly suppressed, and it is justified to expand $U_{\perp}(s)$ by $A_{\perp}(s)$ as in Eq.\eqref{eq:continuous_link}.
Then, a spatial correlation of two link-variables $F(r) \equiv \frac{1}{N_{c}} \langle \mathrm{Tr} U_{x}(0) U^{\dagger}_{x}(ra_{x}) \rangle_{\mathrm{DR}}$ can be written as
\begin{eqnarray}
    \label{eq:UxVx_x}
    F(r)
    & \equiv &
    \frac{1}{N_{c}}
    \langle
    \mathrm{Tr} \; U_{x}(0) U^{\dagger}_{x}(ra_{x})
    \rangle_{\mathrm{DR}} \nonumber \\
    & = &
    \frac{a^{2}}{\beta}
    \langle
    A^{a}_{x}(0) A^{a}_{x}(ra_{x}) 
    \rangle_{\mathrm{DR}}
    +
    \left\{
        1 - \frac{a^{2}}{\beta} \langle A^{a}_{x}(0)^{2} \rangle_{\mathrm{DR}}
    \right\} \nonumber \\
\end{eqnarray}
up to $O(a^{2})$, where $\langle A^{a}_{x}(0)^{2} \rangle_{\mathrm{DR}} = \langle A^{a}_{x}(ra_{x})^{2} \rangle_{\mathrm{DR}}$ 
is used due to the translational symmetry.
The first term is the gluon propagator, and second a constant.
The spatial mass of $A_{x}(s)$ is estimated from the infrared behavior of $F(r)$.

\begin{figure}[h]
    \centering
    \includegraphics[width=8.0cm]{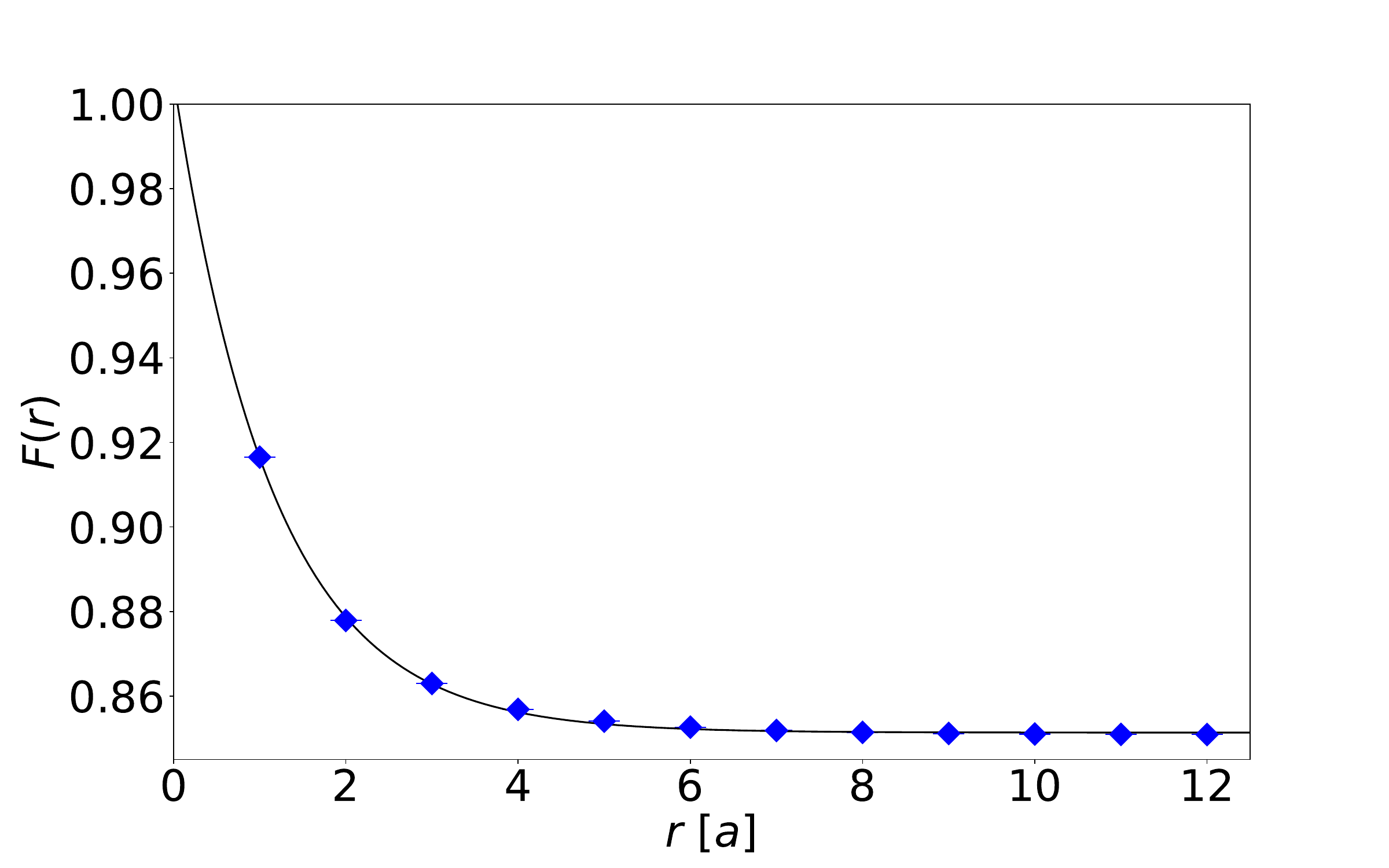}
    \caption{
    The spatial correlation of two $\perp$-directed link-variables, $F(r) \equiv \frac{1}{N_{c}} \langle \mathrm{Tr} ~ U_{x}(0)U^{\dagger}_{x}(ra_{x}) \rangle_{\mathrm{DR}}$, in lattice QCD at $\beta = 6.0$.
    The dots are the lattice data, and the solid line the best exponential fit $A e^{-M_{\perp}r} + B$ with $A \simeq 0.155$, $M_{\perp} \simeq 1.71 \; \mathrm{GeV}$ and $B \simeq 0.851$.
    }
    \label{fig:UxVx_x}
\end{figure}

Figure \ref{fig:UxVx_x} shows the lattice QCD result for $F(r)$ at $\beta = 6.0$, and the lattice QCD data denoted by dots is well reproduced with exponential function
\begin{equation}
    \label{eq:product_UxVx_x}
    F(r)
    \simeq
    A e^{-M_{\perp}r} + B
\end{equation}
with following best fit parameters
\begin{eqnarray}
    \label{eq:F_A}
     A &\simeq& 0.155, \\
    \label{eq:F_M}
     M_{\perp} &\simeq& 0.87a^{-1} \simeq 1.71 \; \mathrm{GeV}, \\
    \label{eq:F_B}
     B &\simeq& 0.851.
\end{eqnarray}
The behavior of the gluon propagator is described by $A$ and $M_{\perp}$, and $B$ corresponds to a constant of the second term in Eq.\eqref{eq:UxVx_x}.
The fact that the value of $B$ is close to unity reflects the strong suppression of the amplitudes of $A_{x}(s)$ and $A_{y}(s)$, as shown in Sec.\ref{subsec:link_distance}.

The spatial mass of $A_{x}(s)$ and $A_{y}(s)$ is estimated as $M_{\perp} \simeq 1.71 \; \mathrm{GeV}$, and thus, they are considered to be massive in the DR gauge.
This result implies that $A_{x}(s)$ and $A_{y}(s)$ are inactive in the infrared region, and then, $A_{t}(s)$ and $A_{z}(s)$ become dominant in the DR gauge.

\subsection{Spatial correlation between two temporal links}
\label{subsec:link_cor}
The results in previous sections imply that, in the DR gauge, 
temporal gluon component $A_{t}(s)$ is dominant for quark confinement 
in the $x$ and $y$ directions.
We here investigate the spatial correlation 
between two temporal link-variables in lattice QCD.

\begin{figure}[h]
    \centering
    \includegraphics[width=6.0cm]{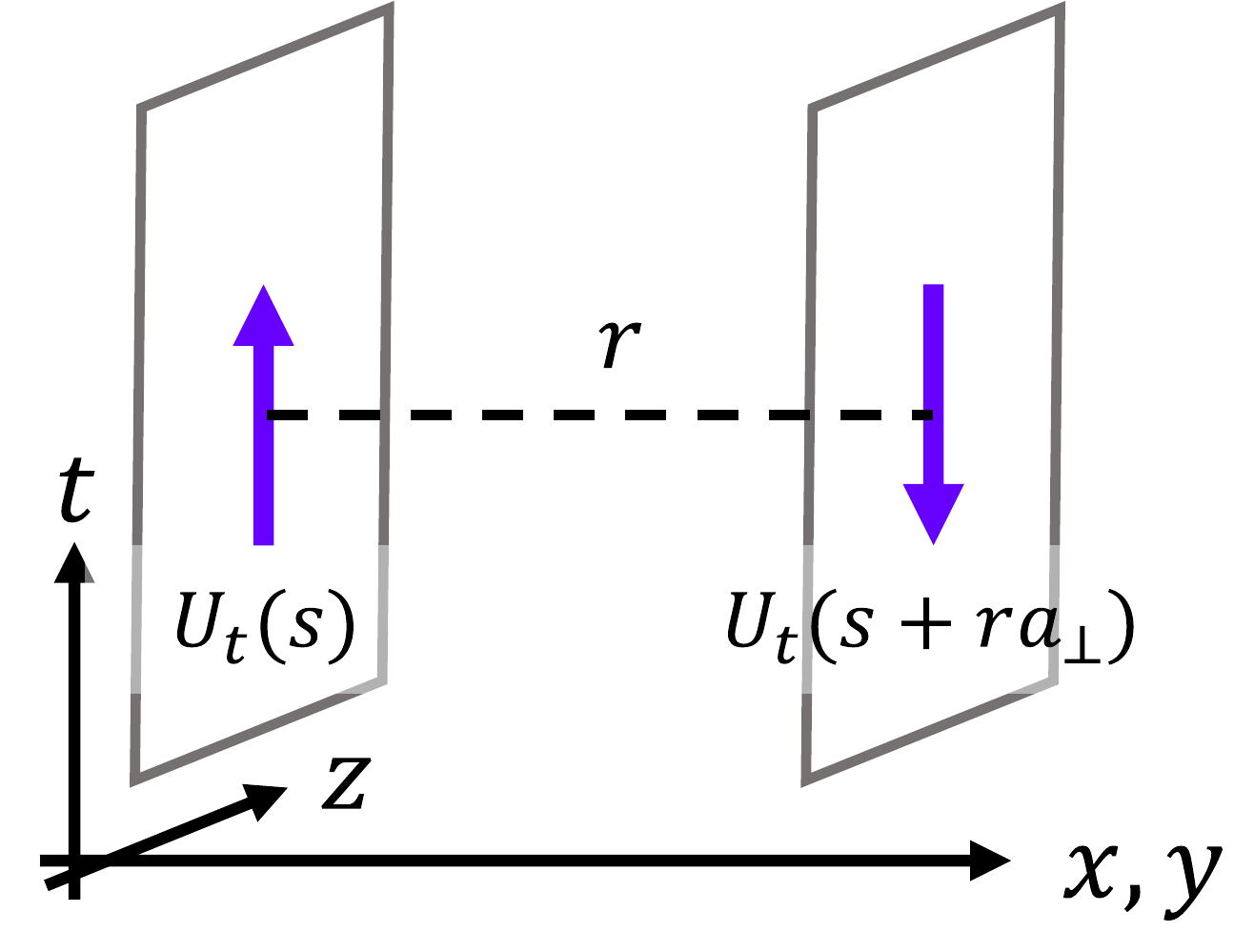}
    \caption{
    Schematic figure of the spatial correlation between link-variables $U_{t}(s)$ and $U^{\dagger}_{t}(s+r a_{\perp})$ on the $t$-$z$ planes separated by $r$ in the $\perp$ direction.
    }
    \label{fig:link_correlation}
\end{figure}

In lattice QCD, the $tz$-projected action is expressed in Eq.\eqref{eq:DRG_lat_act_tzp}, and the local interaction
\begin{equation}
    \label{eq:QCD2_int}
    \beta \sum_{s} 
    \sum_{\mu = t, z}
    \left\{
        1 - \frac{1}{N_{c}}
        \sum_{\perp = x,y}
        \mathrm{ReTr} 
        \left[
            U_{\mu}(s) U^{\dagger}_{\mu}(s+a_{\perp})
        \right]
    \right\}
\end{equation}
provides a distant correlation between $t$-$z$ planes 
in the $x$ and $y$ directions.
Then, 
as shown in Fig.\ref{fig:link_correlation}, 
we calculate the spatial correlation of two temporal link-variables, 
\begin{equation}
    \label{eq:link_correlation}
    C(r)
    \equiv
    \frac{1}{N_{c}}
    \langle 
    \mathrm{ReTr} \;
        U_{t}(s) 
        U^{\dagger}_{t}(s+r a_{\perp})
    \rangle_{\mathrm{DR}},
\end{equation}
in the $\perp$ ($x$ or $y$) direction
with lattice QCD at $\beta = 6.0$. 

\begin{figure}[h]
    \centering
    \includegraphics[width=8.0cm]{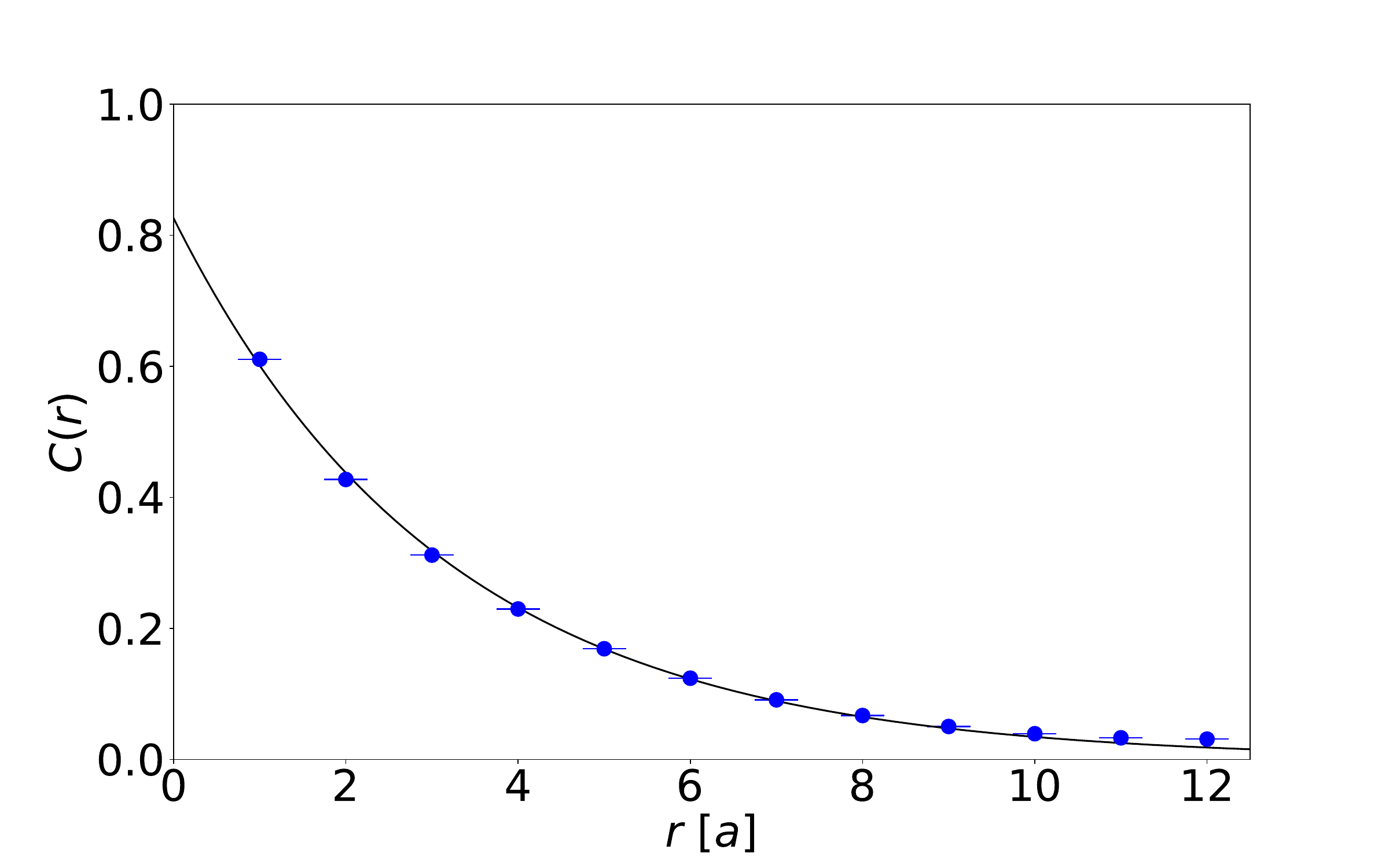}
    \caption{
    The spatial correlation $C(r)$ between two temporal link-variables, 
    $U_{t}(s)$ and $U_{t}(s+ra_{\perp})$, in the DR gauge. 
    The dots are lattice QCD data at $\beta = 6.0$, 
    and the solid line the best exponential fit $Ae^{-mr}$ 
    with $A \simeq 0.83$ and $m \simeq 0.64 ~ \mathrm{GeV}$.
    }
    \label{fig:U4cor}
\end{figure}

Figure \ref{fig:U4cor} shows the lattice QCD result 
of the spatial correlation $C(r)$ of two temporal link-variables, 
plotted against the distance $r$ in the $\perp$ direction. 
The lattice QCD data is well reproduced 
by the exponential function 
\begin{equation}
    \label{eq:fit_curve_exp}
    C(r)
    \simeq
    A e^{-mr} ,
\end{equation}
with the following best fit parameters
\begin{eqnarray}
    \label{eq:A_value}
    && A \simeq 0.83 , \\
    \label{eq:m_value}
    && m \simeq 0.32 a^{-1} \simeq 0.64 \: \mathrm{GeV}.
\end{eqnarray}
Introducing the correlation length $\xi$ defined as
\begin{equation}
    \label{eq:cor_length}
    \xi 
    \equiv
    \frac{1}{m}
    \simeq
    0.31 \; \mathrm{fm},
\end{equation}
the correlation $C(r)$ almost vanishes in larger region than $\xi$.
Thus, the correlation in the $x$ and $y$ directions are short distance and its range is approximately $\xi$.

\section{Discussion}
\label{sec:discussion}

Here, we briefly summarize the previous sections.
In the DR gauge, we have found that the amplitudes of $A_{x}(s)$ and $A_{y}(s)$ are strongly suppressed, and $tz$-projection \eqref{eq:tz_proj} does not affect in quark confinement.
This result implies that two gauge components $A_{t}(s)$ and $A_{z}(s)$ are dominant in the infrared region.
In the DR gauge, the two gauge components $A_{x}(s)$ and $A_{y}(s)$ are found to be massive, and their large mass is conjectured to cause the dominance of $A_{t}(s)$ and $A_{z}(s)$ in the infrared region.
Then, removing the would-be inactive components 
$A_{x}(s)$ and $A_{y}(s)$ in the DR gauge, 
$tz$-projected 4D YM theory is regarded as 
an ensemble of 2D YM-like systems on $t$-$z$ planes 
as shown in Fig.\ref{fig:tz_proj.ed_QCD4_4d}.
These 2D YM-like systems locally interact with neighbors in the $x$ and $y$ directions. Through the interaction, these 2D systems are correlated globally in the $x$ and $y$ directions.
We have also found that the spatial correlation between 2D systems exponentially decreases. 

In this section, we consider a model analysis of 
the $tz$-projected YM theory in the DR gauge, i.e., 
ensemble of 2D YM-like systems on $t$-$z$ planes, which are exponentially correlated within short distance in the $x$ and $y$ directions.

\begin{figure}[h]
    \centering
    \includegraphics[width=6.5cm]{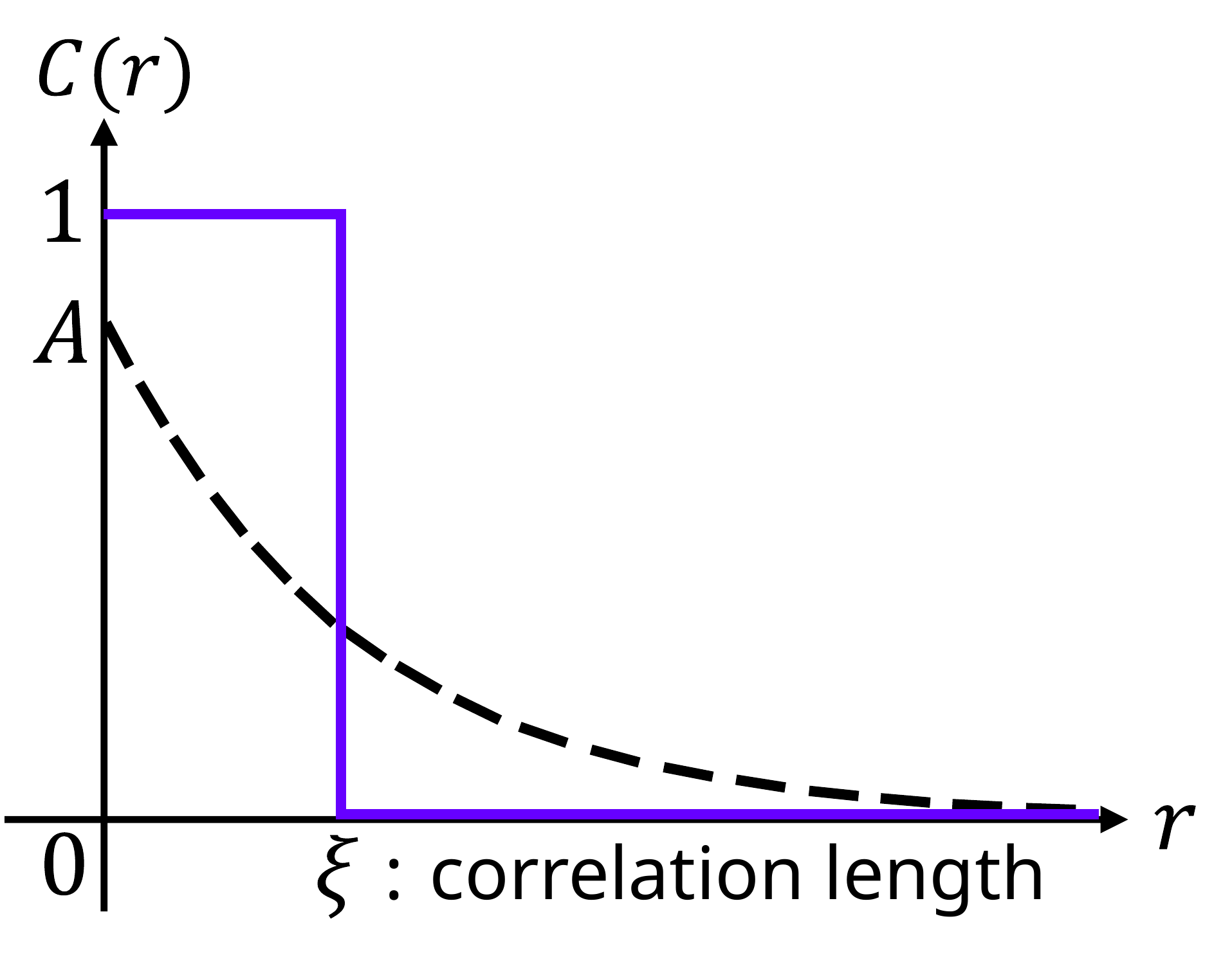}
    \caption{
    A crude approximation of the spatial correlation $C(r)$.
    The broken line represents the exponential correlation, 
    as is obtained in lattice QCD. 
    The solid line is the approximated correlation of $\theta(\xi - r)$ 
    with the correlation length $\xi$.
    }
    \label{fig:step_cor}
\end{figure}

For the analytical modeling, we make a crude approximation of 
replacement of the exponential correlation $C(r)$ 
by a step function, as shown in Fig.\ref{fig:step_cor}, 
\begin{eqnarray}
    \label{eq:step_cor}
    C(r) 
    \to
    \theta(\xi - r)
    =
    \begin{cases}
        1 & (r < \xi) \\
        0 & (r > \xi)
    \end{cases} ,
\end{eqnarray}
where $\xi$ is the correlation length in Eq.\eqref{eq:cor_length}.

Under this approximation, 
when $r$ is smaller than $\xi$, one finds 
$C(r)= \frac{1}{N_{c}} \langle \mathrm{ReTr}~U_{t}(s) U^{\dagger}_{t}(s+ra_{\perp}) \rangle_{\mathrm{DR}} =1$, 
which means $U_{t}(s)=U_{t}(s+ra_{\perp})$. 
In fact, $U_{t}(s)$ can be regarded to be uniform 
within a short distance below $\xi$ in the $x$ and $y$ direction.
The similar relation holds for $U_{z}(s)$ because of symmetry. 

On the other hand, 
when $r$ is larger than $\xi$, one finds 
$C(r)= \frac{1}{N_{c}} \langle \mathrm{ReTr}~U_{t}(s) U^{\dagger}_{t}(s+ra_{\perp}) \rangle_{\mathrm{DR}} =0$, 
and therefore 
the link-variables $U_{t}(s)$ and $U_{t}(s+ra_{\perp})$ 
have no correlation in the $x$ and $y$ directions,  
in other words, their product $U_{t}(s)U_{t}^\dagger (s+ra_{\perp})$ is completely random in the SU($N_c$) manifold.
Also, the similar relation holds for $U_{z}(s)$. 

\begin{figure}[h]
    \centering
    \includegraphics[width=5.0cm]{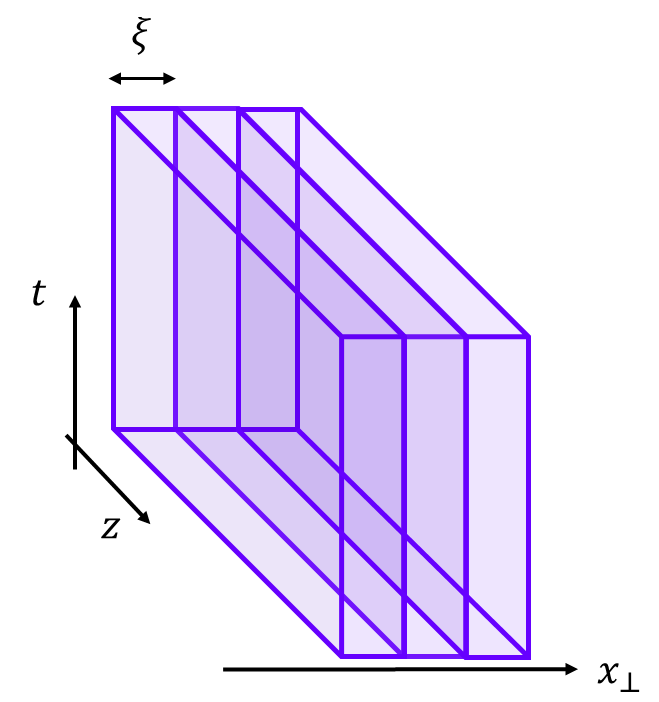}
    \caption{
    The Schematic figure of the DR-gauged YM system under the approximation in Eq.\eqref{eq:step_cor}.
    There is a 2D YM system on each layer, which is independent and do not interact each other.
    }
    \label{fig:tz_layer}
\end{figure}

Therefore, by the approximation \eqref{eq:step_cor}, the $tz$-projected YM theory in the DR gauge can be regarded as an ensemble of 2D YM systems on $t$-$z$ layers, which have the width of $\xi$ and are piled in the $x$ and $y$ direction, as shown in Fig.\ref{fig:tz_layer}.
Within each layer, gluon fields $A_{t}(s)$ and $A_{z}(s)$ are uniform in the $x$ and $y$ direction, and these 2D YM systems are independent and do not interact each other. 

We label these independent layers with two integers $m$ (the $x$-coordinate) and $n$ (the $y$-coordinate).
Then, the gluon fields $A_{\mu}(s)$ on the layer can be expressed as
\begin{equation}
    \label{eq:2d_A}
    A_{\mu}(s)
    =
    A_{\mu}(m\xi, n\xi; t, z)
    \equiv
    A^{M}_{\mu}(t,z)
\end{equation}
in terms of layer index $M=(m, n)$.

Using $A^{M}_{\mu}(t,z)$, the integral over $x$ and $y$ can be replaced by the sum over $m$ and $n$, and the tree-level action  \eqref{eq:DRG_act_tzp} is written as
\begin{equation}
    \label{eq:DRG_QCD_eff_action_IR_2d}
    S^{tz}_{\mathrm{DR}}
    \simeq
    \sum_{N = (m, n)}
    \xi^{2}
    \int dt dz \:
    \mathrm{Tr} \; G^{2}_{tz},
\end{equation}
where the second term in Eq.\eqref{eq:DRG_act_tzp} is dropped off by the approximation \eqref{eq:step_cor}.
This action \eqref{eq:DRG_QCD_eff_action_IR_2d} is described by ``two-dimensional" gluon field $A^{M}_{t}(t,z)$ and $A^{M}_{z}(t,z)$.
Thus, the approximation \eqref{eq:step_cor} reduces 
the DR-gauged YM theory into a ``two-dimensional" theory.

We convert the 4D action \eqref{eq:DRG_QCD_eff_action_IR_2d}
into the corresponding 2D theory,  
by rescaling of the gluon field 
$A^{M}_{\mu}(t,z)$ 
with the correlation length $\xi$ as
\begin{equation}
    \label{eq:xi_A}
    \mathcal{A}^{M}_{\mu}(t,z) 
    \equiv
    \xi A^{M}_{\mu}(t,z).
\end{equation}
Then, the field strength tensor $G^{M}_{\mu \nu}$ and the coupling constant $g$ are also rescaled as
\begin{eqnarray}
    \label{eq:xi_G}
    \mathcal{G}^{M}_{\mu \nu}
    & \equiv &
    \partial_{\mu} \mathcal{A}^{M}_{\nu}
    -
    \partial_{\mu} \mathcal{A}^{M}_{\nu}
    + 
    i \mathfrak{g} 
    \left[
        \mathcal{A}^{M}_{\mu}, \mathcal{A}^{M}_{\nu}
    \right] ,\\
    \label{eq:xi_couple}
    \mathfrak{g}
    & \equiv &
    \frac{g}{\xi}
    =
    m g .
\end{eqnarray}
Using these quantities, the action \eqref{eq:DRG_QCD_eff_action_IR_2d} can be written as
\begin{eqnarray}
    \label{eq:DRG_QCD_eff_action_IR_2d_tree}
    S^{tz}_{\mathrm{DR}}
    & = &
    \int dt dz \:
    \sum_{M}
    \frac{1}{2}
    \mathrm{Tr}
    \left[
        \mathcal{G}^{M}_{\mu \nu}
        \mathcal{G}^{M}_{\mu \nu}
    \right] \nonumber \\
    & = &
    \int dt dz \:
    \frac{1}{4}
    \delta_{M N}
    \mathcal{G}^{a, \,M}_{\mu \nu}
    \mathcal{G}^{a, \,N}_{\mu \nu} ,
\end{eqnarray}
where subscripts $\mu$ and $\nu$ denote $t$ or $z$, and the superscript $a$ the adjoint color index.

According to Eq.\eqref{eq:xi_couple}, the coupling constant $\mathfrak{g}$ acquires a mass dimension through the correlation length $\xi$.
As the general argument in 2D QCD, the coupling constant has a mass dimension, and a scale of the theory must be determined by hand.
However, in the DR-gauged YM theory, the scale is automatically set by the correlation length $\xi$.

From the rescaled action \eqref{eq:DRG_QCD_eff_action_IR_2d_tree}, the interquark potential on the $t$-$z$ plane is calculated as
\begin{equation}
    \label{eq:tree_potential2}
    V_{\mathrm{tree}}(r)
    =
    \frac{\mathfrak{g}^{2}}{2} 
    \frac{4}{3}
    r
\end{equation}
at the tree-level, which 
is a linear potential, i.e., proportional to the interquark distance $r$. 
Using $g = 1.0$ at $\beta = 6/g^{2} = 6.0 $ and $\xi=0.31 \, \mathrm{fm}$, the rescaled coupling is obtained as 
\begin{equation}
    \mathfrak{g}
    \simeq
    0.64 \, \mathrm{GeV}.
\end{equation}
Thus, the interquark potential becomes
\begin{equation}
    \label{eq:tree_potential_value}
    V_{\mathrm{tree}}(r)
    = 
    \sigma_{\mathrm{2D}}
    r,
\end{equation}
with 2D string tension
\begin{equation}
    \label{eq:2d_sigma}
    \sigma_{\mathrm{2D}}
    \simeq 
    1.37 \: \mathrm{GeV/fm},
\end{equation}
which seems to be consistent with the 4D QCD string tension $\sigma \simeq 0.89~\mathrm{GeV/fm}$.

As a caution, this argument is based on a crude approximation \eqref{eq:step_cor}, i.e., the replacement of the exponential correlation by a step function,  
and also this treatment does not include quantum effects.
Then, this value is to be regarded as a rough estimate.
It is however interesting that the estimated 2D string tension 
takes a similar value to 
the string tension of 4D QCD.

\section{Summary and Concluding Remarks}
\label{sec:summary}
In this paper, motivated by one-dimensional color electric flux-tube formation, we have investigated effective dimensional reduction in the 4D YM theory.
We have proposed a new gauge fixing of ``dimensional reduction (DR) gauge'' defined so as to minimize
$R_{\mathrm{DR}} \equiv \int d^{4}s~\mathrm{Tr} \left[ A_{x}^{2}(s) + A_{y}^{2}(s) \right]$, which has a residual gauge symmetry for the gauge function $\Omega(t,z)$ like 2D QCD on the $t$-$z$ plane.
We have defined $tz$-projection as removal of $A_{x,y}(s) \to 0$ in the gauge configurations such as those generated in lattice QCD. 
By the $tz$-projection in the DR gauge, the 4D YM theory is reduced into an ensemble of 2D YM-like systems, which are piled in the $x$ and $y$ directions and interact with neighboring planes.

We have investigated effective dimensional reduction of 4D YM theory in $\mathrm{SU}(3)$ quenched lattice QCD at $\beta = 6.0$.
We have found that, in the DR gauge, the amplitudes of two gauge component $A_{x}(s)$ and $A_{y}(s)$ are strongly suppressed.
In the DR gauge, the interquark potential does not change by the $tz$-projection, and the two components $A_{t}(s)$ and $A_{z}(s)$ play a dominant role in quark confinement. 
For the direction of $\perp=x,y$, we have calculated the spatial correlation $\langle \mathrm{Tr} A_{\perp}(s) A_{\perp}(s+ra_{\perp}) \rangle_{\mathrm{DR}}$ and estimated the spatial mass of $A_{\perp}(s) ~ (\perp = x,y)$ as $M_{\perp} \simeq 1.7 ~ \mathrm{GeV}$ in the DR gauge.
Then, it is conjectured that this large mass makes $A_{\perp}(s)$ inactive and realizes the dominance of $A_{t}(s)$ and $A_{z}(s)$ in infrared region.

We have calculated the spatial correlation of two temporal links, $C(r)\equiv\frac{1}{N_{c}}\langle\mathrm{ReTr}~U_{t}(s)U^{\dagger}_{t}(s+r a_{\perp})\rangle_{\mathrm{DR}}$, and have found that the correlation decreases exponentially as $C(r) \simeq \exp (-mr)$ with $m \simeq 0.6 \; \mathrm{GeV}$, which corresponds to the correlation length $\xi \equiv 1/m \simeq 0.3 \; \mathrm{fm}$. 
According to the dominance of $A_{t}(s)$ and $A_{z}(s)$, we have ignored $A_{x}(s)$ and $A_{y}(s)$ and considered analytical modeling of the 4D YM theory in the DR gauge, using a crude approximation of replacement of $C(r) \to \theta(\xi - r)$ for the spatial correlation.
Under this approximation, the 4D YM theory is found to be regarded as an ensemble of 2D YM systems with the coupling $g_{\mathrm{2D}} \equiv g m$.

In this work, we have used DR gauge fixing 
and have drawn a possible picture of 
effective dimensional reduction in the 4D YM theory.
To be strict, however, the DR gauge fixing has 
the Gribov ambiguity \cite{Gribov_1978}, 
which is a general problem appearing along with most gauge fixing 
such as the Landau, Coulomb and MA gauges.
To confirm the obtained picture, 
it is desired to perform more careful checks on 
effects from the Gribov ambiguity.

Finally, we list below future works related to this subject. 
It is necessary to perform the lattice QCD calculations for various $\beta$ and to investigate the scaling property 
in the DR gauge.
Also, it is desired to improve the crude approximation of the spatial correlation, $C(r)\to\theta(\xi-r)$, in Sec.~\ref{sec:discussion}.
For example, the correlation $C(r)$ can be represented by multi step functions as $C(r) \simeq \sum_{i}C(\frac{\xi_{i}+\xi_{i+1}}{2})\theta(\xi_{i+1}-r)\theta(r-\xi_{i})$ with appropriate $\xi_{i}$.
By taking a more appropriate approximation, 
more realistic correspondence might be obtained 
between the 4D YM theory and 2D systems. 

To include quark degrees of freedom is also an important 
future work. In particular, spontaneous breaking of chiral symmetry is a typical non-perturbative property, and it would be valuable to investigate the chiral condensate $\langle \Bar{q}q \rangle$ in 4D DR-gauged QCD after the $tz$-projection.

It is also interesting to examine 
the behavior of dynamical quarks in 4D DR-gauged QCD.
As shown in Sec.\ref{subsec:transverse_mass} and Sec.\ref{subsec:link_cor}, gluon propagation in the $x$ and $y$ directions is suppressed.
According to this, it is considered that gluons are bounded on $t$-$z$ planes in DR-gauged QCD.
However, quarks are not expected to be bounded on the $t$-$z$ plane, because the realization of a real 2D system 
contradicts spontaneous chiral symmetry breaking 
which is realized in 4D QCD, due to the Coleman theorem. 
Thus, DR-gauged QCD system would be a system in which gluons (bosons) are bounded on $t$-$z$ planes and quarks (fermions) propagate between the planes.
This system is similar to the graphene \cite{graphene}, 
where electrons (fermions) are bounded on 2D planes and photons (bosons) propagate between planes.
Thus, the QCD system in the DR gauge might be regarded as the ``dual'' graphene, where roles of fermion and boson are interchanged.

Considering 4D DR-gauged QCD at finite temperatures is another future work.
At finite temperatures, a linear potential disappears, and the Coulomb or Yukawa potential between (anti)quarks is realized.
As temperatures increase, the dimensional reduction picture is considered to be broken.
It seems interesting to investigate how the dimensional reduction picture changes and breaks down at high temperatures.

In the DR gauge, $A_{t}(s)$ and $A_{z}(s)$ are considered to be strongly correlated and propagate over long distances in the $t$ and $z$ directions, like the 2D YM theory on the $t$-$z$ plane.
As a while, the spatial correlation $C(r)$ of two temporal link-variables decreases exponentially as shown in Sec.~\ref{subsec:link_cor}, and this means that the propagation of $A_{t}(s)$ and $A_{z}(s)$ in the $x$ and $y$ directions is suppressed in the DR gauge.
Thus, in the DR gauge, $A_{t}(s)$ and $A_{z}(s)$ seem to have anisotropic masses.
This property seems to suggest a similarity between 4D DR-gauged QCD and a fracton system \cite{fracton}, 
where propagation of an quasi-particle excitation is restricted 
in some direction. 

We have investigated effective dimensional reduction of 
the 4D YM theory in the DR gauge, and using a crude approximation, we have described the 4D YM system 
in terms of 2D gauge degrees of freedom.
This suggests a possibility that an essence of 4D QCD can be expressed with two-dimensional degrees of freedom.
In other words, there is a possibility that 4D QCD is a holograph which is constructed from a hologram of the essential 2D field variables, which might lead an idea of ``hologram QCD''.

\section*{Acknowledgement}
\label{sec:acknowledgement}
H.S. was supported in part by a Grants-in-Aid for
Scientific Research [19K03869] from Japan Society for the Promotion of Science. 
The lattice QCD calculations have been performed by SQUID at Osaka University.
\vspace{0.5cm}

\appendix

\section{$xy$-projected Wilson loop in DR gauge}
\label{sec:xy_proj_Wlp}
In Appendix \ref{sec:xy_proj_Wlp}, we consider $xy$-projected Wilson loop in the DR gauge and derive Eq.\eqref{eq:res_gt_xy_wilsonloop3}.
Since the Wilson lines $L_{\mathrm{upper}}$ and $L_{\mathrm{bottom}}$ are elements of $\mathrm{SU}(N_{c})$, they can be expressed as
\begin{eqnarray}
    \label{eq:L_upper_expand}
    && L_{\mathrm{upper}}
    = 
    L^{0}_{\mathrm{upper}} 1
    +
    L^{a}_{\mathrm{upper}} T^{a}, \\
    \label{eq:L_bottom_expand}
    && L_{\mathrm{bottom}}
    = 
    L^{0}_{\mathrm{bottom}} 1
    +
    L^{a}_{\mathrm{bottom}} T^{a},
\end{eqnarray}
where $L^{0}_{\mathrm{upper}}, L^{a}_{\mathrm{upper}}, L^{0}_{\mathrm{bottom}} ~\text{and}~ L^{a}_{\mathrm{bottom}}$ are generally complex numbers.

Using the notation in Sec.\ref{subsec:WilsonLoop_DR_proj.ed}, the $xy$-projected Wilson loop transforms as
\begin{eqnarray}
    \label{eq:res_gt_xy_wilsonloop2}
    W^{xy}(r,T)
    \to 
    && \mathrm{Tr}
    \left[
        \tilde{\Omega} 
        \left(
            L^{0*}_{\mathrm{upper}} 1
            +
            L^{a*}_{\mathrm{upper}} T^{a}
        \right)
    \right. \nonumber \\
    && \hspace{16pt}
    \left.
        \tilde{\Omega}^{\dagger}
        \left(
            L^{0}_{\mathrm{bottom}} 1
            +
            L^{b}_{\mathrm{bottom}} T^{b}
        \right)
    \right] \nonumber \\
    = &&
    N_{c}
    L^{0*}_{\mathrm{upper}}
    L^{0}_{\mathrm{bottom}} 
    \nonumber \\
    && +
    \mathrm{Tr}
    \left[
        \tilde{\Omega} 
        L^{a*}_{\mathrm{upper}} T^{a}
        \tilde{\Omega}^{\dagger}
        L^{a}_{\mathrm{bottom}} T^{a}
    \right],
\end{eqnarray}
where $\mathrm{Tr} \; T^{a} = 0$ is used.
While the second term is not gauge-invariant for the residual gauge transformation with gauge function $\Omega(t,z)$, 
the first term is invariant. 
For the vacuum expectation value of the $xy$-projected Wilson loop, only the invariant term survives as
\begin{eqnarray}
    \label{eq:res_gt_xy_wilsonloop3_after}
    \langle 
        W^{xy}(r,T)
    \rangle_{\mathrm{DR}}
    & = &
    N_{c}
    \left\langle
    L^{0*}_{\mathrm{upper}}
    L^{0}_{\mathrm{bottom}}
    \right\rangle_{\mathrm{DR}} \nonumber \\
    & = &
    \frac{1}{N_{c}}
    \left\langle
        \mathrm{Tr} \; L^{\dagger}_{\mathrm{upper}}
        \mathrm{Tr} \; L_{\mathrm{bottom}}
    \right\rangle_{\mathrm{DR}}.
\end{eqnarray}
Thus, the $xy$-projected Wilson loop $\langle W^{xy}(r,T) \rangle_{\mathrm{DR}}$ is generally non-zero and finite.

\section{Tree-level potential in 2D QCD}
\label{sec:tree_2D}
In Appendix \ref{sec:tree_2D}, we derive the tree-level interquark potential \eqref{eq:tree_potential2} in 2D QCD on the $t$-$z$ plane, using the same notation in Sec.\ref{sec:discussion}.
In Euclidean spacetime, the generating functional of 2D QCD at the quenched level is written as
\begin{eqnarray}
    \label{eq:Z_2d_tree}
    Z^{\mathrm{2D}}
    & = &
    \int \mathcal{D A} \;
    \exp
    \left[
        - \int dt dz \;
    \right.
        \left(
            \frac{1}{4}
            \delta_{M N}
            \mathcal{G}^{a, \,M}_{\mu \nu}
            \mathcal{G}^{a, \,N}_{\mu \nu} 
        \right. \nonumber \\
          && \hspace{110pt} 
          + 
          J^{a, \, M}_{\mu} \mathcal{A}^{a, \, M}_{\mu}
        \Bigg)
    \Bigg] \nonumber \\
    & = &
    \int \mathcal{D A} \:
    \exp
    \left[
        - \int dt dz \:
        \left(
            \frac{1}{4}
            \mathcal{A}^{a, \, M}_{\mu}
            D^{ab}_{\mu \nu}
            \delta_{M N}
            \mathcal{A}^{b, \, N}_{\mu}
        \right.
    \right. \nonumber \\
    && \hspace{95pt}
            +
            J^{a, \, M}_{\mu} \mathcal{A}^{a, \, M}_{\mu}
            +
            \cdots
        \Bigg)
    \Bigg], \\
    \label{eq:D_2d}
    D^{ab}_{\mu \nu}
    & \equiv &
    \left(
        - \partial^{2} \delta_{\mu \nu} + \partial_{\mu} \partial_{\nu}
    \right)
    \delta^{ab} ,
\end{eqnarray}
where $J^{a, \, M}_{\mu}$ is the color current coupling to the gauge field as $J^{a, \, M}_{\mu} \mathcal{A}^{a, \, M}_{\mu}$, and $(\cdots)$ contains the third and fourth orders of $\mathcal{A}^{a, \, M}_{\mu}$.
Completing the square and performing the Gaussian integral for $\mathcal{A}^{a, \, M}_{\mu}$, the generating functional \eqref{eq:Z_2d_tree} becomes
\begin{eqnarray}
    \label{eq:Z_2d_tree_current}
    Z^{\mathrm{2D}}
    && \simeq
    \exp
    \left[
        - \frac{1}{2}
        \int d^{2} \zeta 
        \int d^{2} \zeta'
    \right. \nonumber \\
    && \hspace{35pt}
    \left.
        J^{a, \, M}_{\mu} (\zeta)
        \left(
            D^{-1}
        \right)^{ab}_{\mu \nu}
        \delta_{M N}
        J^{b, \, N}_{\nu} (\zeta') 
    \right],
\end{eqnarray}
where $\zeta = (t,z)$ denotes the two-dimensional coordinate and we have ignore the higher orders of $\mathcal{A}^{a, \, M}_{\mu}$ to consider at tree-level.
The operator $\left(D^{-1}\right)^{ab}_{\mu \nu}$ is the inverse of $D^{ab}_{\mu \nu}$.

Using the conservation law of the color current
\begin{equation}
    \label{eq:conserve_current}
    \partial_{\mu} J^{a, \, M}_{\mu}
    =
    0,
\end{equation}
the generating functional \eqref{eq:Z_2d_tree_current} is expressed as
\begin{eqnarray}
    \label{eq:Z_2d_tree_current2}
    \hspace{-20pt}
    Z^{\mathrm{2D}}
    && \propto
    \exp
    \left[
        - \frac{1}{2}
        \int d^{2} \zeta 
        \int d^{2} \zeta'
    \right. \nonumber \\
    && 
    \left.
        J^{a, \, M}_{\mu} (\zeta)
        \left(
            \int \frac{d^{2} p}{(2\pi)^{2}}
            \frac{\delta_{\mu \nu}}{p^{2}} e^{i p (\zeta - \zeta')}
        \right)
        J^{a, \, M}_{\nu} (\zeta') 
    \right].
\end{eqnarray}
Considering a static quark at $z=z_{1}$ and a static antiquark at $z=z_{2}$, the color current is expressed as
\begin{equation}
    \label{eq:static_current}
    J^{a, \, M}_{\mu}
    =
    \mathfrak{g}
    \left[
        \delta(z-z_{1}) - \delta(z-z_{2})
    \right]
    \delta_{\mu 0}
    \delta^{M M_{0}}
    T^{a},
\end{equation}
where $M_{0}$ is a layer index of the layer in which the quark and the antiquark stay.
Substituting this into Eq.\eqref{eq:Z_2d_tree_current2}, the generating function is calculated as
\begin{eqnarray}
    \label{eq:Z_2d_source_act}
    Z^{\mathrm{2D}}
    \propto
    \exp
    &&
    \left[
        - \frac{\mathfrak{g}^{2}}{2} C_{2} \delta^{M_{0} N_{0}}
    \right. \nonumber \\
    &&
    \left.
        \int_{-\infty}^{\infty} dt \:
        \left(
            \int_{-\infty}^{\infty} \frac{d p}{\pi}
            \frac{1}{p^{2}} 
            (1 - \cos pr)
        \right)
    \right],
\end{eqnarray}
where $C_{2}$ is the Casimir operator of SU($N_{c}$) in the fundamental representation and $r \equiv |z_{1}-z_{2}|$.
The $\delta^{N_{0} M_{0}}$ means that no interaction works between particles on different layers in the approximation of Eq.\eqref{eq:step_cor}.
From the generating functional, the effective potential $V_{\mathrm{eff}}$ is derived as
\begin{equation}
    \label{eq:Z_pot}
    Z^{\mathrm{2D}}
    \sim
    \exp
    \left[
        - \int dt~V_{\mathrm{eff}}
    \right],
\end{equation}
and the tree-level potential is obtained as a linear potential,
\begin{eqnarray}
    \label{eq:tree_potential}
    V_{\mathrm{tree}}(r)
    & = &
    \frac{\mathfrak{g}^{2}}{2} 
    C_{2} \:
    \delta^{M_{0} N_{0}}
    \int_{-\infty}^{\infty} \frac{d p}{\pi}
    \frac{1}{p^{2}} 
    (1 - \cos pr ) \nonumber \\
    & = &
    \frac{\mathfrak{g}^{2}}{2} 
    \frac{4}{3}
    \delta^{M_{0} N_{0}}
    r.
\end{eqnarray}

\section{Interquark potential in $\perp$ direction with crude approximation}
\label{sec:perp_potential}
In Appendix \ref{sec:perp_potential}, we consider the $\perp$-directed interquark potential in the DR gauge with the crude approximation shown in Sec.\ref{sec:discussion}.
The potential is calculated from the $tz$-projected Wilson loop on the $t$-$\perp$ plane.
Under the approximation \eqref{eq:step_cor}, if two Wilson lines $L_{\mathrm{left}}$ and $L_{\mathrm{right}}$ are in the same layer, they are identical
\begin{equation}
    L_{\mathrm{left}}
    =
    L_{\mathrm{right}}
\end{equation}
because $C(r) = 1$ means $U_{t}(s) = U_{t}(s+ra_{\perp})$.
Then, the $tz$-projected Wilson loop on the $t$-$\perp$ plane is calculated as
\begin{equation}
    W^{tz}(r,T)
    =
    \mathrm{Tr}
    \left[
        L^{\dagger}_{\mathrm{left}} L_{\mathrm{right}}
    \right]
    =
    \mathrm{Tr}[1].
\end{equation}

On the other hand, if $L_{\mathrm{left}}$ and $L_{\mathrm{right}}$ are not in the same layer, they have no correlation and their product $L^{\dagger}_{\mathrm{left}}L_{\mathrm{right}}$ is completely random.
In this case, the vacuum expectation value of the $tz$-projected Wilson loop on the $t$-$\perp$ plane becomes zero
\begin{equation}
    \langle 
        W^{tz}(r,T)
    \rangle_{\mathrm{DR}}
    =
    \left\langle
        \mathrm{Tr}
        \left[
            L^{\dagger}_{\mathrm{left}} L_{\mathrm{right}}
        \right]
    \right\rangle_{\mathrm{DR}}
    =
    0.
\end{equation}
Thus, under the approximation \eqref{eq:step_cor}, the Wilson loop on the $t$-$\perp$ plane is simplified to be
\begin{equation}
    \label{eq:step_t_perp_wlp}
    \langle 
        W^{tz}(r,T)
    \rangle_{\mathrm{DR}}
    =
    \begin{cases}
        \mathrm{Tr}[1] & (r < \xi) \\
        0 & (r >  \xi)
    \end{cases},
\end{equation}
and the interquark potential in the $\perp$ direction is calculated as
\begin{equation}
    \label{eq:step_t_perp_potential}
    V(r) 
    = 
    - \!
    \lim_{T \to \infty}
    \frac{1}{T} \ln \left\langle W^{tz}(r,T) \right\rangle_{\mathrm{DR}} = 
    \begin{cases}
        0 & (r < \xi) \\
        \infty & (r > \xi)
    \end{cases}.
\end{equation}
Therefore, the approximation \eqref{eq:step_cor} leads to quark confinement with the infinite string tension in $x$ and $y$ directions.

In reality, the interquark potential in the $\perp$ direction is expected to be milder than that in Eq.\eqref{eq:step_t_perp_potential}, since the spatial correlation $C(r)$ does not decreases $\theta$-functional, but exponentially.

\end{document}